\documentclass[12pt]{article}
\usepackage{amsmath,amssymb,epsfig,arydshln,graphicx,rotating}
\usepackage{color}
\input{colordvi.tex}
\def\unit{{\relax{\rm 1\kern-.26em I}}}

\newcommand{\lsim}{\lesssim}
\newcommand{\gsim}{\gtrsim}
\newcommand{\tr}{{\rm Tr}}

\def\6#1{{\underline{#1}}}
\def\m6#1{{\underline{#1}\,}}

\newdimen\Tdim
\def\ispan{{\setbox0=\hbox{i}%
\Tdim\ht0\advance\Tdim\dp0\rule[-\dp0]{0pt}{\Tdim}}}
\def\jspan{{\setbox0=\hbox{j}%
\Tdim\ht0\advance\Tdim\dp0\rule[-\dp0]{0pt}{\Tdim}}}
\def\Tspan#1{{\setbox0=\hbox{#1}%
\Tdim\ht0\advance\Tdim\dp0\advance\Tdim.55ex\rule[-\dp0]{0pt}{\Tdim}\box0}}

\def\be{\begin{eqnarray}}
\def\ben{\begin{eqnarray*}}
\def\ee{\end{eqnarray}}
\def\een{\end{eqnarray*}}

\def\p{\partial}

\def\D{\mathcal{D}}

\def\=:{=\hspace{-.7em}\raisebox{1.1ex}{.}\hspace{.1em}\raisebox{-0.2ex}{.} }

\newcommand {\beq}{\begin{eqnarray}}
\newcommand {\eeq}{\end{eqnarray}}
\newcommand {\non}{\nonumber\\}

\makeatletter

\renewcommand\section{\@startsection {section}{1}{\z@}%
                                   {-3.5ex \@plus -1ex \@minus -.2ex}%
                                   {2.3ex \@plus.2ex}%
                                   {\normalfont\large\bfseries}}

\renewcommand\subsection{\@startsection{subsection}{2}{\z@}%
                                     {-3.25ex\@plus -1ex \@minus -.2ex}%
                                     {1.5ex \@plus .2ex}%
                                     {\normalfont\normalsize\bfseries}}

\newcount\hour \newcount\minute
\hour=\time \divide \hour by 60
\minute=\time
\def\now{%
\ifnum \hour<13
  \ifnum \hour=0 \advance \hour by 12 \number\hour:\else \number\hour:\fi%
     \ifnum \minute<10 0\fi%
     \number\minute%
\ A.M.%
\else \advance \hour by -12 \number\hour:%
  \ifnum \minute<10 0\fi%
  \number\minute%
  \ P.M.%
\fi%
}

\makeatother

\begin{document}

\baselineskip=18pt  
\numberwithin{equation}{section}  
\allowdisplaybreaks  



%
%


\thispagestyle{empty}

\vspace*{-2cm}
\begin{flushright}
{\tt YGHP-16-05}
\end{flushright}

\begin{flushright}
\end{flushright}

\begin{center}

\vspace{-.5cm}

\vspace{0.5cm}
{\bf\Large Stabilizing semilocal strings by polarization}

\vspace{0.5cm}
Minoru Eto${}^{1}$,
Muneto Nitta${}^{2}$ and
Kohei Sakurai${}^{1}$

\vspace{0.5cm}
{\it\small 
${}^1$ Department of Physics, Yamagata University, Yamagata, 990-8560, Japan}\\
{\it\small
${}^2$ Department of Physics, and Research and Education 
Center for Natural Sciences,}\\  
{\it\small 
 Keio University, Hiyoshi 4-1-1, Yokohama, Kanagawa 223-8521, Japan}

\vspace{1cm}
{\bf Abstract}\\

\end{center}

Semilocal strings are vortices in the extended Abelian-Higgs model 
with two complex Higgs scalar fields among which a global $SU(2)$ symmetry acts. 
They are known to be stable (unstable against expansion)  
in type-I (II) superconductors, in which 
gauge field is heavier (lighter) than the Higgs scalar field.
In this paper,
we find that vortices can be stabilized in the whole parameter region 
including the type-II region  
by adding a potential term breaking the $SU(2)$ symmetry.
We construct numerical solutions in various parameters 
and determine the vortex phase diagram consisting of six phases.
In two phases, 
a vortex is polarized, that is, 
split into two half-quantized vortices with a certain distance,
to form a vortex molecule, 
while in the rests a vortex is identical to 
the conventional Abrikosov-Nielsen-Olesen vortex.

\vspace{0.5cm}
\parbox{15cm}{
\small\hspace{15pt}
}


\clearpage
\setcounter{page}{1}
\setcounter{footnote}{0}
\renewcommand{\thefootnote}{\arabic{footnote}}

\section{Introduction}

Magnetic fields are confined  in the form of vortices or flux tubes 
in superconductors, 
macroscopically described 
by the Abelian-Higgs model, that is, 
a $U(1)$ gauge theory coupled with 
one complex Higgs scalar field \cite{Abrikosov:1956sx,Nielsen:1973cs}.
Vortices are called Abrikosov-Nielsen-Olesen (ANO) vortices or local vortices.
Depending on the masses $m_e$ and $m_H$ of 
gauge and Higgs fields, respectively, 
the superconductor can be classified into type-I ($m_H <m_e$) or II ($m_e < m_H$).
At the critical coupling ($m_e=m_H$), it is called 
Bogomol'nyi-Prasado-Sommerfield (BPS).
For type-I, II and BPS superconductors,
there exist attractive, repulsive, and no forces, respectively, 
between vortices.
Superconductors are stable against applied magnetic fields 
when they are of type-II, constituting a vortex lattice inside it 
stabilized by repulsion among vortices. 
Vortices are cosmic strings in cosmology, and so 
relativistic dynamics have been studied well. 

Semilocal cosmic strings are vortex strings 
in the extended Abelian-Higgs model with two Higgs complex scalar fields 
with an $SU(2)$ symmetry
\cite{Vachaspati:1991dz,Achucarro:1999it}.
Cosmological consequences of semilocal strings such as 
their effects on cosmic microwave background were 
studied in Ref.~\cite{Urrestilla:2007sf}.
Semilocal cosmic strings 
reduce to $O(3)$ sigma model (${\mathbb C}P^1$ model) 
lumps in strong gauge coupling limit \cite{Hindmarsh:1991jq},
 which are supported by the second homotopy group 
 $\pi_2(M)$ and have size and phase moduli. 
 The stability of semilocal strings were studied very well \cite{Hindmarsh:1991jq,Achucarro:1992hs}.
As lumps, semilocal strings at critical coupling (BPS limit)
have size and phase moduli and are marginally stable.
At near critical coupling a potential is induced for 
the size modulus.
In the type-II region in which the gauge boson is lighter than the Higgs boson, 
semilocal strings are unstable to expand, 
while they shrink to the ANO vortices and are stable in the type-I region in which 
the Higgs boson is lighter than the gauge boson. 
Non-Abelian semilocal cosmic strings have been studied recently 
\cite{Shifman:2006kd,Eto:2007yv},
which reduce to Grassmann sigma model lumps 
in strong gauge coupling limit.
Reconnection of two colliding non-Abelin semilocal strings were also studied \cite{Eto:2006db}.
Other than semilocal strings, possible semilocal solitons were classified 
\cite{Gibbons:1992gt,Hindmarsh:1992yy},
including codimension-four sigma model instantons 
\cite{Hindmarsh:1992ef,Eto:2015obh}.

\bigskip
On the other hand, 
the topological charges of topological solitons 
supported by certain homotopy groups 
are usually quantized to be integers.
However in certain situations, the minimum topological charge can be fractional.
Typical examples are
fractional magnetic vortices (flux tubes) in 
multi-component or multi-band superconductors
\cite{Babaev:2001hv,Babaev:2004rm,Smiseth:2004na,Gurevich:2003,Goryo:2007,Nitta:2010yf,fractional-exp} 
and 
fractional superfluid vortices in two-component
\cite{Mueller:2002,Kasamatsu:2003,Son:2001td,Kasamatsu:2004,Kasamatsu:2005,Eto:2011wp,Aftalion:2012,Cipriani:2013nya,Kobayashi:2013wra},
three-componet \cite{Eto:2012rc,Cipriani:2013wia},
and multi-component 
\cite{Eto:2013spa,Nitta:2013eaa} Bose-Einstein condensates (BECs).
For the system of $N$ components, 
topological charge is fractionalized to be $1/N$ 
if VEVs are all equal.
Each fractional vortex is a half-local and half-global vortex 
in superconductors while it is a global vortex in BECs.
$O(3)$ sigma model (${\mathbb C}P^1$ model) lumps in 2+1 dimensions 
(or instanton in 2+0 dimensions) 
are characterized by $\pi_2(M)$.
A lump can be decomposed into 
a vortex anti-vortex pair with fractional lump charges 
in the presence of a certain potential term 
\cite{Schroers:1995he,Schroers:1996zy,Nitta:2011um,Alonso-Izquierdo:2014cza}.
Topologically the same thing occurs for baby Skyrmions 
characterized by $\pi_2(M)$. 
One baby Skyrmion is decomposed into 
a vortex anti-vortex pair with fractional lump charges 
in the presence of the same type of the potential term
\cite{Jaykka:2010bq,Kobayashi:2013aja,Kobayashi:2013wra}.
As a 3+1 dimensional example,
a Skyrmion characterized by the third homotopy group $\pi_3(M)$ 
can be decomposed into a monopole and anti-monopole pair 
with fractional Skyrmion charges in the presence of a certain potential term 
\cite{Gudnason:2015nxa}.

In this paper, we decompose a semilocal string  
into two half-quantized strings by introducing a potential term 
breaking the $SU(2)$ symmetry
and show that semilocal strings become 
stable against expansion in the whole parameter region including 
type-II superconductors.
Each fractional string has
opposite charges $\pm 1/2$ of a global $U(1)$ symmetry 
of a subgroup of $SU(2)$, which is unbroken by 
the additional potential.
We obtain numerical solutions for the fractional strings, and also investigate the asymptotic behaviors
quite different from either the well-known ANO strings or usual semilocal strings.
We find that the asymptotic behaviors of the profile function 
decay exponentially with the smallest masses of the fields at the bulk. 
We study the dependence of the polarization of the single semilocal string 
in all parameter region 
with the masses $m_e$, $m_\lambda$ and $m_\eta$ in detail.
We find that the semilocal string is stable in the whole parameter region. 
Especially, the type II region $m_e < m_\lambda$ is divided into
two phases. 
In one region, the unpolarized semilocal string, namely the type-II ANO solution, appears for $m_e < m_\lambda < m_e + m_\eta$.
In the other region $m_\lambda > m_e + m_\eta$, the two Higgs fields have zeros at different points, namely the string is polarized.
The displacement $|d|$ of the two zeros is larger for smaller $m_e$. It also increases as $m_\lambda$ increase but saturates at a certain upper
value.

The behaviors of asymptotics of the profile functions,  
that they exponentially decay with the smallest mass in the bulk, 
are similar 
\cite{Eto:2009kg,Eto:2013hoa} 
to those of the semisuperfulid non-Abelian strings 
\cite{Balachandran:2005ev,Nakano:2007dr}
in the dense QCD, 
even though the roles of local and global symmetry
are opposite to the model studied in this paper 
(namely,  the Abelian symmetry is global and the non-Abelian symmetry is local).
Thus, we believe that the phenomena that the smallest mass controls asymptotic behavior is common to strings 
in a wide range of physical model.

This paper is organized as follows.
In Sec.~\ref{sec:SU2}, we review semilocal vortices 
in the $SU(2)$ symmetric model.
In Sec.~\ref{sec:polarization}, we work  out 
the vortex structure in the model with a broken $SU(2)$ symmetry.
We construct numerical solutions with various parameters 
and 
determine the phase diagram.
Section \ref{sec:summary} is devoted to a summary and discussion.
The numerical method that we use in this paper is explained in Appendix \ref{sec:NR}.

\newpage
\section{The semilocal strings in an $SU(2)$ symmetric model}\label{sec:SU2}

We consider the Abelian-Higgs model with a Higgs doublet $H = (H_1,H_2)$. The Lagrangian is given by
\be
{\cal L} = - \frac{1}{4e^2}F_{\mu\nu}F^{\mu\nu} + \D_\mu H(\D^\mu H)^\dagger - V_0,\quad
V_0 = \frac{\lambda^2}{2}\left(HH^\dagger - v^2\right)^2,
\label{eq:lag}
\ee
with $F_{\mu\nu} = \p_\mu A_\nu - \p_\nu A_\mu$ and $\D_\mu H = (\p_\mu + i A_\mu) H$.
The Lagrangian has a global $SU(2)$ symmetry 
under which the Higgs fields is transformed as the fundamental representation.  
In the vacua, both the gauge and flavor symmetry are spontaneously broken. 
The vacuum states are degenerate and form a vacuum manifold $S^{3}$ defined by
\be
HH^\dagger = v^2.
\ee
By identifying each $U(1)$ gauge orbit as a point, the vacuum moduli is given by
\be
{\cal M}_0 = \frac{SU(2)}{U(1)} \simeq \mathbb{C}P^{1}.
\ee
Therefore, the number of the physical Nambu-Goldstone modes is $\dim_{\mathbb{R}}{\cal M}_0 = 2$, and
there is a massive Higgs mode with mass
\be
m_\lambda = \sqrt{2}\lambda v.
\ee 
At the same time, the photon gets  mass
\be
m_e = \sqrt{2}ev.
\ee

It is well-known that the model (\ref{eq:lag}) admits solitonic strings, the so-called semilocal strings,  even though
the vacuum manifold is homotopically trivial $\pi_1(S^{3}) = 1$.
It is a static solution of the classical equation of motion
\be
\D_\mu \D^\mu H_a &=& \left[\lambda^2 (v^2 - HH^\dagger) \right]\! H_a,\quad(a=1,2),
\label{eq:eom0_1}\\
\frac{1}{e^2} \p_\mu F^{\mu\nu} &=& - i \left(H\D^\nu H^\dagger - \D^\nu H H^\dagger\right).
\label{eq:eom0_2}
\ee
It has been known that the semilocal string is dynamically  stable only when
the masses satisfy the relation for type-I superconductors
\be
m_\lambda \le m_e.
\label{eq:mass_cond}
\ee

At the critical coupling $e=\lambda$ ($m_\lambda = m_e$), the equations of motion reduce to a set of the first order differential 
equations, so-called Bogomol'nyi equation,
\be
(\D_1 + i \D_2) H = 0,\qquad
\frac{1}{e^2} F_{12} = HH^\dagger - v^2.
\ee
Introducing the complex coordinate $z = x^1+ix^2$ and $\p_z = (\p_1 - i \p_2)/2$, the first equation 
can be solved by \cite{Eto:2005yh,Eto:2006pg}
\be
H = v e^{- \frac{\psi(z,\bar z)}{2}}H_0(z),\quad A_1 + i A_2 = -2i\p_{\bar z} \psi,
\ee
where $H_0(z)$ is a $2$-vector whose components are holomorphic functions of $z$ 
and $\psi$ is a complex scalar function of $z$ and $\bar z$ determined below. 
In the following, we will set $\psi$ to be real positive by 
fixing the $U(1)$ gauge degree of freedom. 
Plugging these into the second Bogomol'nyi equation, we end up with the master equation for the 
semilocal vortex
\be
\frac{1}{2e^2v^2} \p_i^2 \psi = 1 - H_0H_0^\dagger e^{-\psi},
\ee
with $\p_i^2 = \p_1^2 + \p_2^2$ and $F_{12} = - \frac{1}{2}\p_i^2 \psi$, and the boundary condition
\be
\psi \to \log H_0H_0^\dagger \quad  \text{as} \quad |z| \to \infty.
\ee
The master equation determines $\psi$ according to a given $H_0(z)$. $H_0$ is a pair of two polynomials $P_1(z)$
and $P_2(z)$ as  
\be
H_0(z) = \left(P_1(z),\ P_2(z)\right).
\ee
The tension (mass per unit length) of the semilocal strings at the critical coupling is determined only by the quantized 
magnetic flux $2\pi k$ with an integer $k$ as
\be
T =  2\pi v^2 |k|.
\ee
The integer $k$ is related to the highest degree of the polynomials in $H_0(z)$. Indeed, $\psi$ asymptotically behaves as
$\psi \to \log |z|^{2k} + \cdots$, then we find
\be
-\frac{1}{2\pi}\int dx^1dx^2\ F_{12} = \frac{1}{4\pi}\int dx^1dx^2\ \p_i^2 \psi = \frac{1}{4\pi}\oint_{S^1_\infty} dS\ \vec n \cdot \vec \nabla \psi = k.
\ee
The minimal BPS semilocal string is characterized by the following $H_0$ with 2 complex parameters 
\be
H_0 = \left(z-z_1,\ z-z_2\right),\quad \{z_a\} \in \mathbb{C}^{2},
\label{eq:mm_lambdal}
\ee
where we have chosen the {\it symmetric} boundary configuration $H \to (v,\ v)/\sqrt{2}$ up to the overall $U(1)$ phase.
Since $H$ is proportional to $H_0$, the $a$-th Higgs field becomes zero at $z = z_a$, and so one might expect $2$ peaks in energy density for $k=1$ configuration, but it is not the case. 
What we observe is only a single peak. Namely, the energy distribution is always axially symmetric,
and no substructures can be found. 

\begin{figure}
\begin{center}
\includegraphics[width=16cm]{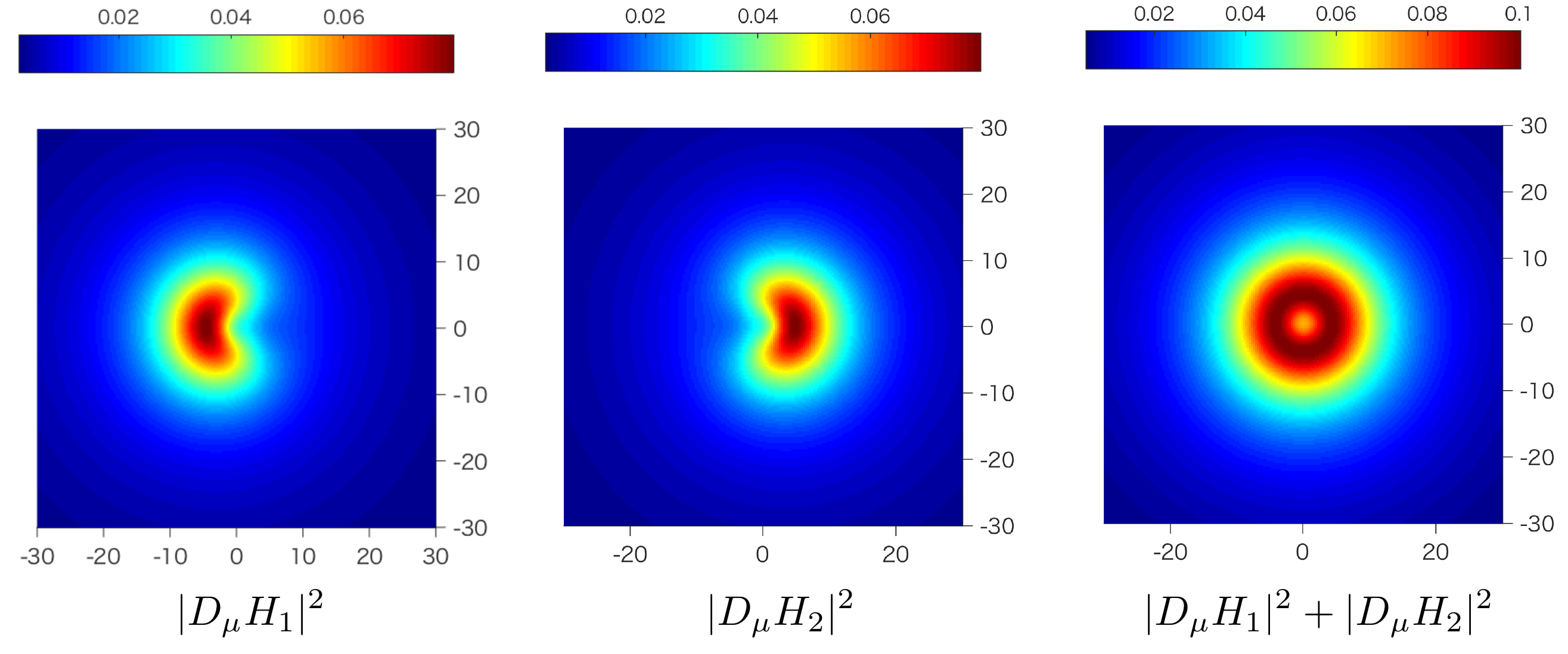}
\caption{The scalar kinetic energies for the single BPS semilocal string with $H_0 = (z-5,\ z+5)$ in $N_{\rm F} =2$ case. }
\label{fig:BPS}
\end{center}
\end{figure}

As an example, we show the kinetic energy densities $|\D_\mu H_1|^2$, $|\D_\mu H_2|^2$
and $|\D_\mu H_1|^2 + |\D_\mu H_2|^2$ for  $H_0 = (z-5,\ z+5)$ in Fig.~\ref{fig:BPS}.
Nevertheless the individual kinetic energies have peaks at apparently different points, the sum  $|\D_\mu H_1|^2 + |\D_\mu H_2|^2$
is axially symmetric and has only one peak at the origin. The gauge kinetic term and scalar potential term are also axially symmetric,
so no inner structures appear. 
This may easily be understood by looking at another solution which is just obtained by rotating $H_0$ 
given in Eq.~(\ref{eq:mm_lambdal}) with a certain $SU(2)$ flavor transformation to the following one
\be
H_0 = \left(z-z_1,\ z-z_2\right) \quad \to \quad \sqrt{2}\,\left(z-\frac{z_1+z_{2}}{2},\ \frac{z_1-z_2}{2}\right).
\label{eq:zero_shift}
\ee
It is now clear that only the first Higgs field vanishes at the center of mass $z=\frac{z_1+z_2}{2}$ while the second component never touches zero.
Thus, it is quite natural to regard $(z_1 + z_2)/2$ as the position of the semilocal string, indeed the energy density has a single peak there.
The parameter $(z_1 - z_2)/2$ in the second component of $H_0$ 
can be decomposed into the phase modulus and 
the ``distance'' $|z_1-z_2|$ which represents 
the thickness (size) of the semilocal string. 
When $|z_1 - z_2|$ is zero,
the second component in the Higgs field plays no role, namely is everywhere zero, so that the semilocal string becomes the ANO string
in the Abelian-Higgs model with a single complex field. The semilocal string for $|z_1 - z_2| > 0$ is a fatter string.
In the opposite limit $|z_1 - z_2| \to \infty$, the first component is negligible compared to the second one. This means that the semilocal string
gets  fat and dilutes as $|z_1-z_2|$ being increased, and finally disappears with the vacuum left behind.

\begin{figure}[ht]
\begin{center}
\includegraphics[height=6cm]{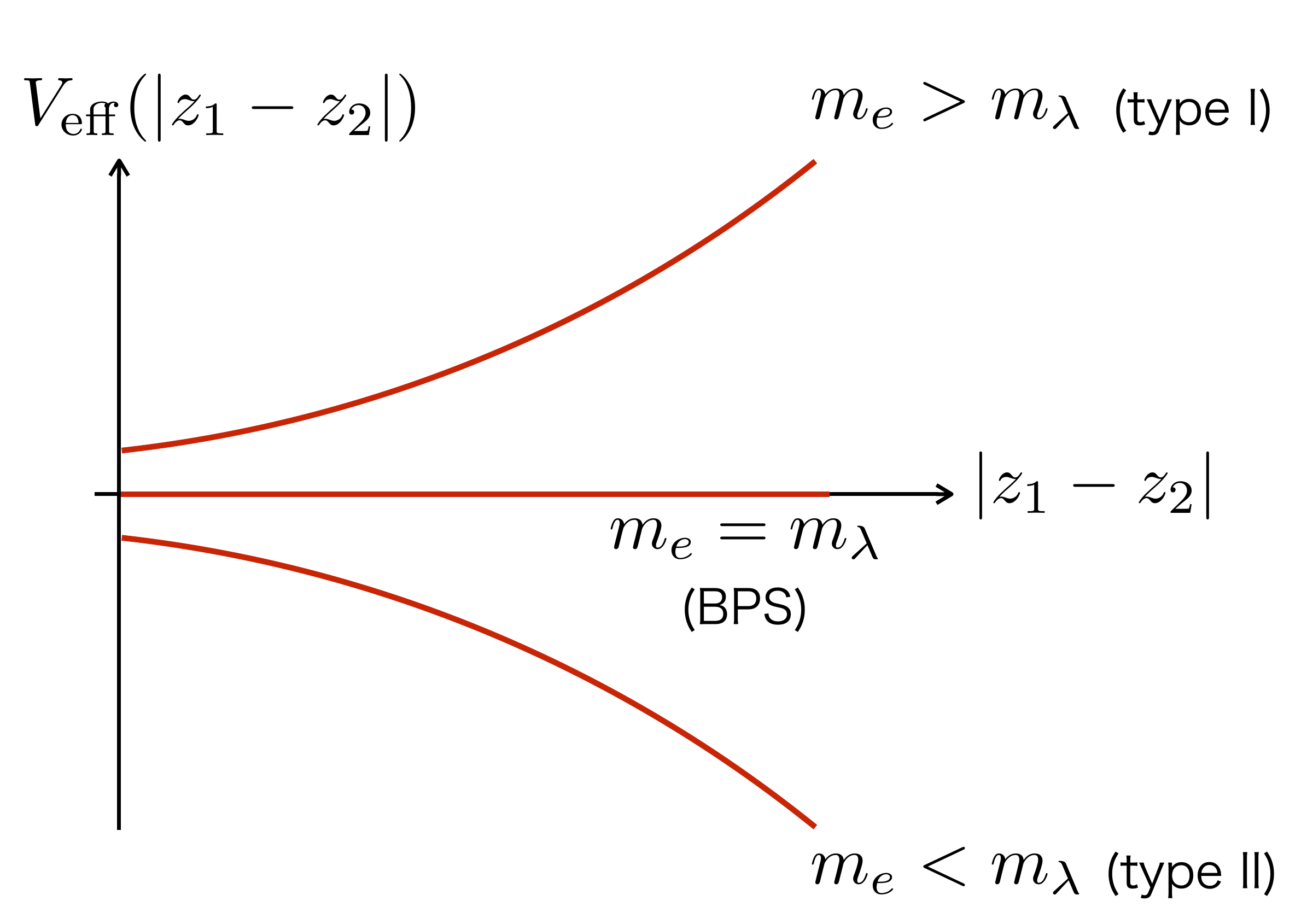}
\caption{A schematic image for an effective potential of the size moduli $|z_1-z_2|$.}
\label{fig:effpot_size_moduli}
\end{center}
\end{figure}

Once we leave from the critical coupling $m_e = m_\lambda$, the semilocal string is no longer BPS. As a consequence, all the moduli fields 
except for the center of mass are gone.
Accordingly, the moduli $z_1-z_2$ of the BPS semilocal string is lifted by an effective potential.
A schematic image of the effective potential is shown in Fig.~\ref{fig:effpot_size_moduli}. 
It falls down to zero for $m_\lambda < m_e$ (type I) 
or runs away to infinity for $m_e < m_\lambda$ (type II).
In other words, the semilocal string in $m_\lambda < m_e$ is 
the same as the ANO string of the type I, while one in $m_e < m_\lambda$ is unstable to dilute.

\section{Polarization of semilocal strings}\label{sec:polarization}

\subsection{The $SU(2)$ breaking interaction}

In this section, we will introduce polarization for the semilocal string.
As we will explain below, the polarization cannot be manifestly defined for the semilocal strings in the $SU(2)$ symmetric model reviewed in
the previous section. It will turn out that a key ingredient for the polarization is breaking of $SU(2)$ symmetry.
For that purpose, we will include an additional
Higgs potential to the Lagrangian ${\cal L}$ given in Eq.~(\ref{eq:lag}),
\beq
V_1 = \frac{\eta^2}{2} (H\tau_3 H^\dagger)^2 
= \frac{\eta^2}{2} (|H_1|^2 - |H_2|^2)^2,
\label{eq:add_pot}
\eeq
with $\tau_3$ being the third element of Pauli matrix.
The flavor symmetry $SU(2)$ is explicitly  broken
to the $U(1)$ subgroup generated by $\tau_3$.\footnote{
Note that the additional Higgs potential (\ref{eq:add_pot}) is identical 
to the $D$-term potential in the context of the supersymmetric extension 
of our model when we gauge the $U(1)$ symmetry generated by $\tau_3$.
}

\begin{figure}[t]
\begin{center}
\includegraphics[height=6cm]{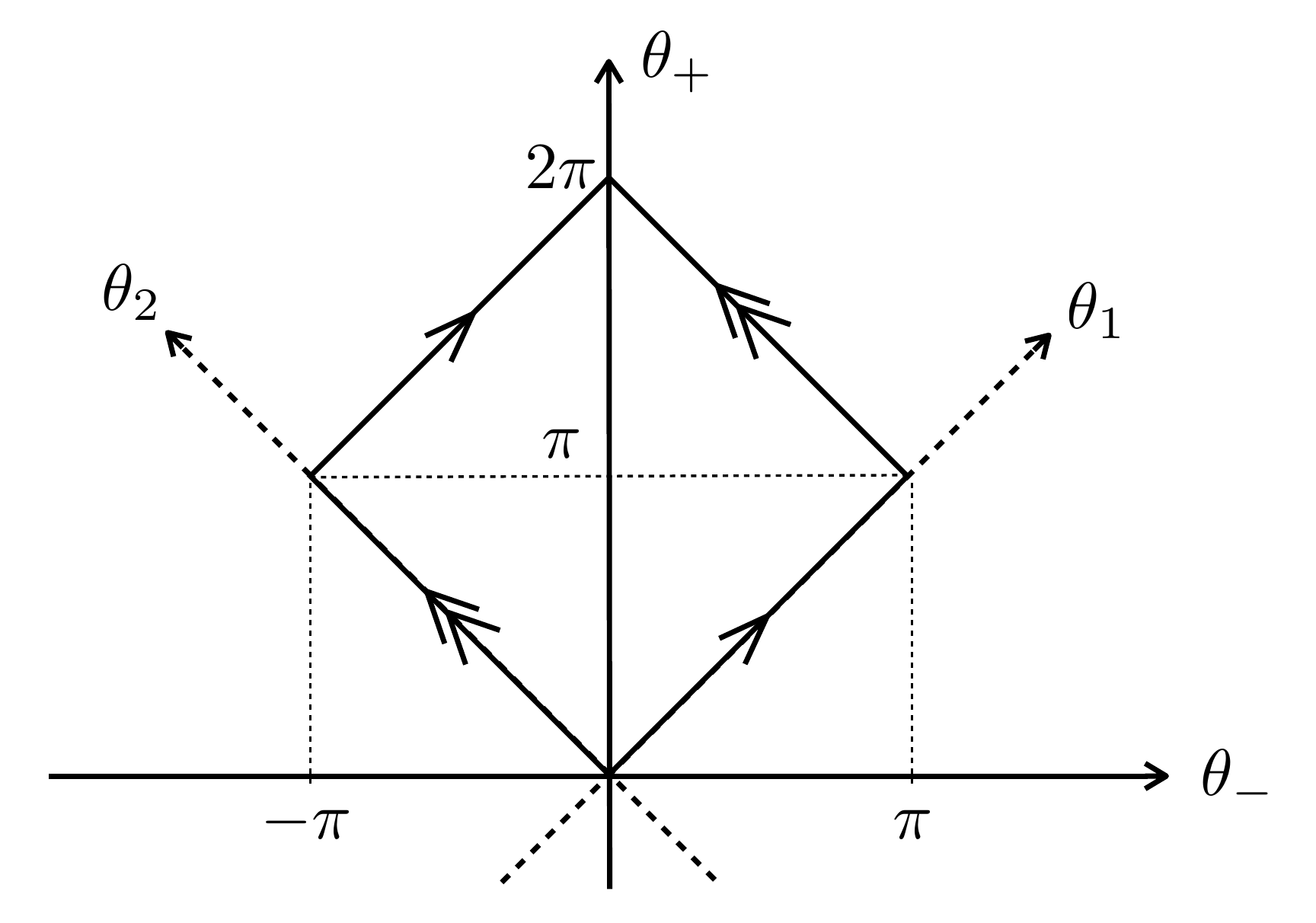}
\caption{Vacuum manifold}
\label{fig:torus}
\end{center}
\end{figure}

The additional potential reduces the original vacuum manifold $S^3$ ($|H_1|^2 + |H_2|^2 = v^2$) to $S^1 \times S^1$ defined by
the following condition
\beq
|H_1|^2 = |H_2|^2 = \frac{v^2}{2}. 
\label{eq:vac_cond_mod}
\eeq
The vacuum manifold is parametrized by the phases of $H_1$ and $H_2$ as 
\beq
H_1 = \frac{v}{\sqrt 2} e^{i\theta_1} = \frac{v}{\sqrt 2}e^{i\theta_+ + i\theta_-} ,\quad 
H_2 = \frac{v}{\sqrt 2} e^{i\theta_2} = \frac{v}{\sqrt 2}e^{i\theta_+ - i\theta_-}.
\eeq
The gauge orbit is parametrized by $\theta_+$ while the global orbit is parametrized by $\theta_-$.
The vacuum manifold is illustrated in Fig.~\ref{fig:torus}.

To see the mass spectra, let us consider small fluctuations around the vacuum
$H_1 = \frac{v}{\sqrt{2}} + f_1 + i g_1$ and $H_2 = \frac{v}{\sqrt 2} + f_2 + i g_2$. The quadratic Lagrangian
is given by
\be
{\cal L}^{(2)} &=& - \frac{1}{4e^2}\tilde F_{\mu\nu}^2+ \left(\p_\mu \frac{f_1+f_2}{\sqrt2}\right)^2 + \left(\p_\mu \frac{f_1-f_2}{\sqrt2}\right)^2 
+ \left(\p_\mu \frac{g_1 - g_2}{\sqrt{2}}\right)^2\non
&& +~ v^2 \tilde A_\mu^2 - 2\lambda^2v^2\left(\frac{f_1+f_2}{\sqrt2}\right)^2 - 2\eta ^2v^2\left(\frac{f_1-f_2}{\sqrt2}\right)^2,
\label{eq:quadratic_lag}
\ee
where we have defined $\tilde A_\mu = A_\mu + v^{-1}\p_\mu \frac{g_1+g_2}{\sqrt 2}$.
\begin{table}[h]
\begin{center}
\caption{Mass spectrum around the vacuum $H_1 = H_2 = v/\sqrt{2}$.}
\ \\
\begin{tabular}{c||c|c|c|c}
fields & NG & gauge & Higgs 1 & Higgs 2\\
\hline
mass & $0$ & $m_e=\sqrt{2}\, ev$ & $m_\lambda = \sqrt{2}\, \lambda v$ & $m_\eta = \sqrt{2}\, \eta v$
\end{tabular}
\end{center}
Thus the masses of the gauge and Higgs fields remain intact. There exist two Nambu-Goldstone modes for $\eta=0$. 
Now,  one of the Nambu-Goldstone mode, $\chi = \frac{f_1-f_2}{\sqrt2}$,
gets mass $m_\eta = \sqrt{2} v \eta$, while the other mode associated with the spontaneously broken relative $U(1)$ symmetry remains massless.
\end{table}
\begin{figure}[ht]
\begin{center}
\includegraphics[width=8cm]{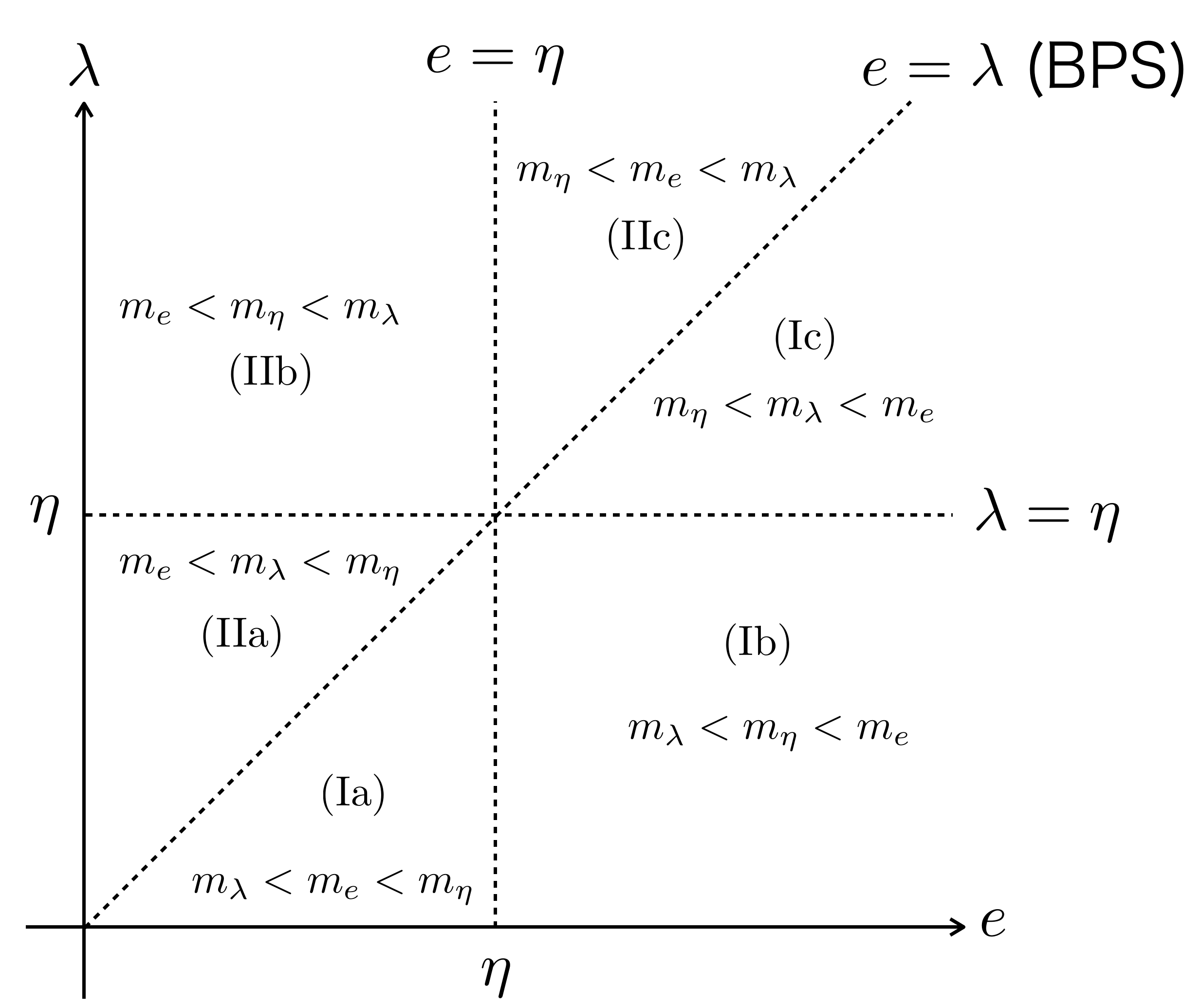}
\caption{Classification of the parameter space.}
\label{fig:elambda}
\end{center}
\end{figure}

For convenience, let us classify the parameter space $(e,\lambda,\eta)$ into six regions: We refer regions to the type I if $m_s < m_e$  while
to the type II if $m_e < m_\lambda$, as denoted above. 
According to $m_\eta$,
each region is further divided into three classes 
which we denote by a, b and c from large $m_\eta$ to small $m_\eta$,
as shown Fig.~\ref{fig:elambda}.

The additional potential $V_1$ in Eq.~(\ref{eq:add_pot}) changes the equation of motion for the Higgs fields (\ref{eq:eom0_1})
as 
\be
\D_\mu \D^\mu H_1 &=& \left[\lambda^2 (v^2 - HH^\dagger) 
- \eta^2\left(|H_1|^2 - |H_2|^2\right)\right] H_1 ,
\label{eq:eom0_1_mod}\\
\D_\mu \D^\mu H_2 &=& \left[\lambda^2 (v^2 - HH^\dagger) 
- \eta^2\left(|H_2|^2 - |H_1|^2\right)\right] H_2.
\label{eq:eom0_2_mod}
\ee
The equations of motion for the gauge field are given in Eq.~(\ref{eq:eom0_2}).
The boundary conditions for the Higgs fields at spacial infinity is that in Eq.~(\ref{eq:vac_cond_mod}). 
That for the gauge fields will be given below.

\subsection{String solutions}

\subsubsection{General arguments}

In what follows, we will concentrate on straight string solutions. For that purpose, we assume the configuration is static, $\p_0 = 0$, and
also we impose $\p_3 = 0$, namely, the string is parallel to the $x^3$ axis.

Let us first concentrate on asymptotic behaviors of the fields.
Those for the Higgs fields are given by
\beq
H_1 \to \frac{v}{\sqrt 2} e^{ik_1\theta} = \frac{v}{\sqrt 2} e^{i\frac{k_1 + k_2}{2}\theta} e^{i\frac{k_1 - k_2}{2}\theta},\quad
H_2 \to \frac{v}{\sqrt 2} e^{ik_2\theta} = \frac{v}{\sqrt 2} e^{i\frac{k_1 + k_2}{2}\theta} e^{-i\frac{k_1 - k_2}{2}\theta},
\label{eq:asymp}
\eeq
with $k_1$ and $k_2$ being arbitrary integers, and
$r = \sqrt{x^2 + y^2}$ and $\tan \theta = y/x$.
Similarly, the asymptotic form of the gauge field is given by
\beq
A_i \to -\alpha\p_i \theta = \alpha \epsilon_{ij} \frac{x^j}{r^2}.
\eeq
Here, $\alpha$ is  
the total magnetic flux 
\beq
\alpha = -\frac{1}{2\pi}\int d^2x\ F_{12}.
\label{eq:magnetic_flux}
\eeq
Then, the kinetic energy of the Higgs field reads 
\beq
{\cal K}_H = \sum_{a=1}^2 |D_i H_a|^2 \to \left(\frac{v^2}{2}\sum_{a=1}^2(k_a-\alpha)^2\right) r^{-2}.
\eeq
Therefore, the tension (mass per unit length) of the string solution is 
\beq
K_H = \int d^2x\ {\cal K}_H = 2\pi \left(\frac{v^2}{2}\sum_{a=1}^2(k_a-\alpha)^2\right) \int^\Lambda \frac{dr}{r} = \pi v^2 \sum_{a=1}^2(k_a-\alpha)^2 \log \Lambda + \cdots,
\eeq
where $\Lambda$ is an IR cutoff (or the size of the system).
This is minimized when
\beq
\alpha = \frac{k_1 + k_2}{2},\quad 
K_H = \frac{\pi v^2}{2}\left(k_1 - k_2\right)^2 \log \Lambda + \cdots.
\label{eq:alpha}
\eeq

For finite energy configuration (in infinite space $\Lambda \to \infty$), 
we consider a common integer as $k \equiv k_1 = k_2$. This leads $\alpha = k$, and the string tension becomes finite since the logarithmic divergent
term disappears. 
Namely, we impose the usual finite energy condition
\beq
D_\mu H_a = \p_\mu H_a + i A_\mu H_a \to 0,\quad (a=1,2)
\eeq
at spatial infinity.
When we go around the string, both the phases $\theta_1$ and $\theta_2$ of the Higgs fields rotates by $2\pi k_a$ as
\beq
\frac{1}{2\pi} \oint \vec\nabla \theta_a \cdot d\vec r = k_a.
\eeq
Rephrasing this in terms
of $\theta_\pm$, the gauged phase changes by $2\pi k$ while the global phase $\theta_-$ 
is constant as 
\beq
\frac{1}{2\pi} \oint \vec\nabla \theta_\pm \cdot d\vec r = \frac{1}{2\pi} \oint \vec\nabla \frac{\theta_1\pm\theta_2}{2} \cdot d\vec r = \frac{k_1\pm k_2}{2} \equiv k_\pm.
\eeq
We will refer $k_+$ to the local charge and $k_-$ to the global charge.
Namely, this string solution with $(k_1,k_2) = (1,1)$, or $[k_+,k_-] = [1,0]$, is purely a local vortex.
In the following sections, we will investigate the local string in detail.

Let us describe the generic axially symmetric solution in more details. We make the following ansatz
\beq
H_a = \frac{v}{\sqrt{2}}e^{ik_a\theta}F_a(r),\quad
A_i = \frac{k_1+k_2}{2} \epsilon_{ij}\frac{x^j}{r^2} A(r).
\eeq
The equations of motion for $F_a(r)$ and $A(r)$ read
\beq
F_1'' + \frac{F_1'}{r} - \left(\frac{k_1 - \frac{k_1+k_2}{2}A}{r}\right)^2 F_1 + \frac{m_\lambda^2}{2}\left(1-\frac{F_1^2+F_2^2}{2}\right)F_1 - \frac{m_\eta^2}{4}(F_1^2-F_2^2)F_1 = 0,\label{eq:eom_axial1}\\
F_2'' + \frac{F_2'}{r} - \left(\frac{k_2 - \frac{k_1+k_2}{2}A}{r}\right)^2 F_2 + \frac{m_\lambda^2}{2}\left(1-\frac{F_1^2+F_2^2}{2}\right)F_2 - \frac{m_\eta^2}{4}(F_2^2-F_1^2)F_2 = 0,\label{eq:eom_axial2}\\
(k_1+k_2)\left(A'' - \frac{A'}{r}\right) + m_e^2 \left[\left(k_1 - \frac{k_1+k_2}{2}A\right)F_1^2 + \left(k_2 - \frac{k_1+k_2}{2}A\right)F_2^2\right] = 0.
\label{eq:eom_axial3}
\eeq

\subsubsection{Fractional string}

Here we consider a global vortex with a fractional flux (local charge)
which is not allowed in the $SU(2)$ symmetric model with $\eta=0$.
The simplest example is $(k_1,k_2) = (1,0)$ for which 
only $H_1$ has a nontrivial winding. From Eq.~(\ref{eq:alpha}), we should choose $\alpha = \frac{1}{2}$. The corresponding string tension diverges as
\beq
K_H\left(\alpha = \frac{1}{2}\right)  = \frac{\pi v^2}{2}\log \Lambda + \cdots.
\label{eq:half_global}
\eeq
The logarithmic divergence is universal property for the global strings. 
Note that the tension of $n$ Abelian global strings is known as $T = 2\pi v^2 n^2 + \cdots$.
This formula with Eq.~(\ref{eq:half_global}) implies that out solution with $(k_1,k_2) = (1,0)$ is the global string with a half winding number $n= \frac{1}{2}$.
Similarly, the magnetic flux for this string is $-\pi$ which is a half of that for the integer ANO string as expected from Eq.~(\ref{eq:magnetic_flux}). 
These charges can be understood from the local/global charge
of this solution $[k_+,k_-] = \left[\frac{1}{2},\frac{1}{2}\right]$.

\begin{figure}[t]
\begin{center}
\includegraphics[width=16cm]{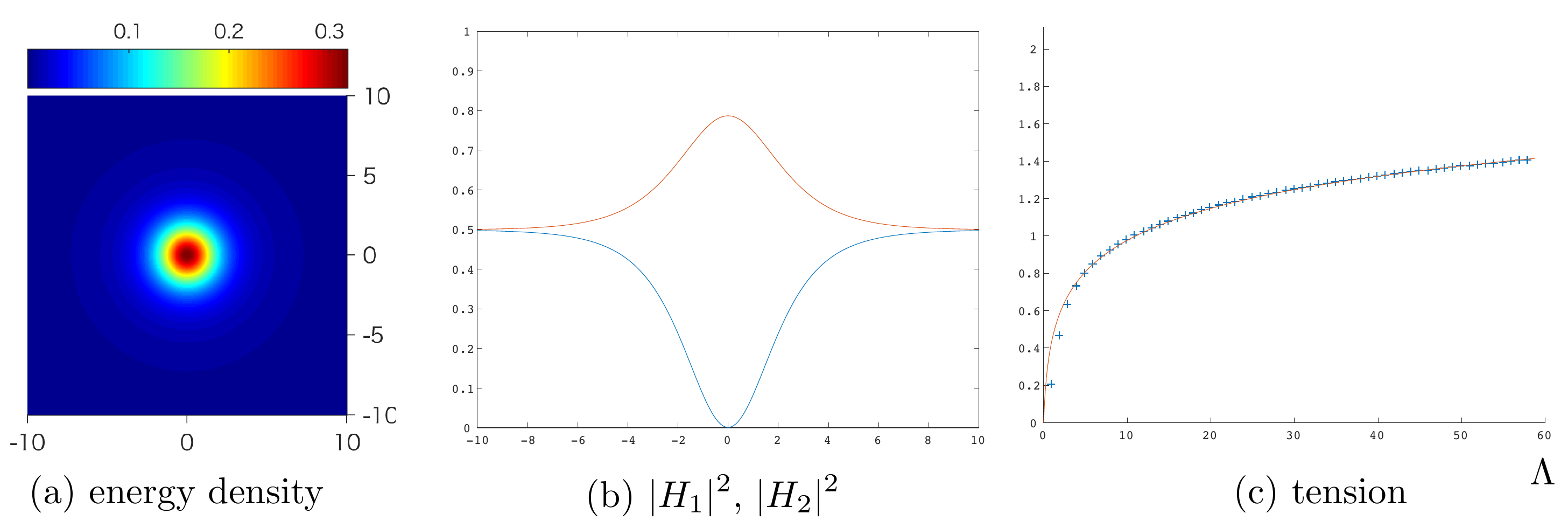}
\caption{$(k_1,k_2) = (1,0)$ string solution for $(e,\lambda,\eta) = (0.3, 1, 0.25)$.
The panel (a) shows the energy density in the $xy$ plane, and (b) shows the profile functions $|H_1|^2$ (blue) and $|H_2|^2$ (red).
The tension as a function of the IR cutoff $\Lambda$ is shown in the panel (c) where
the red curve and blue crosses correspond to the analytical expectation curve in Eq.~(\ref{eq:expected_curve})
and numerical data, respectively.}
\label{fig:local_global_vortex}
\end{center}
\end{figure}

The minimal string with $(k_1,k_2) = (1,0)$ is axially symmetric. So we solve Eqs.~(\ref{eq:eom_axial1}) -- (\ref{eq:eom_axial2})
with the following boundary conditions
\beq
F_1(0) = 0,\quad F_2'(0) = 0,\quad A(0) = 0,\qquad
F_1(\infty) = 1,\quad F_2(\infty) = 1,\quad A(\infty) = 1.
\eeq
A numerical solution 
is shown in Fig.~\ref{fig:local_global_vortex}. 
The profile function of the winding component $H_1$ vanishes 
at the center of the string core, where that of the
unwinding component $H_2$ has non-zero expectation value. 
The profile functions are shown in the middle panel of Fig.~\ref{fig:local_global_vortex}.
Since this string has a half global string charge, its tension should diverge as explained in Eq.~(\ref{eq:half_global}).
We compute the tension $E(\lambda) = 2\pi \int^\Lambda_0 dr\ r H$ as a function of the IR cutoff $\Lambda$ and compare it with
analytical expectation formula
\beq
\frac{E(\Lambda)}{2\pi v^2} = \frac{1}{4}\log \Lambda + {\rm const.}\,,
\label{eq:expected_curve}
\eeq
in the panel (c) of Fig.~\ref{fig:local_global_vortex}. They agree quite well. We also numerically integrate $F_{12}/{2\pi}$ and get $-0.5000$.
Thus, our solution indeed have fractional local and global charge $1/2$ and $1/2$.

Next, let us examine asymptotic behavior of the solution by perturbing the fields around the background solution as
\beq
F_1 = 1 - \frac{f_+(r)+f_-(r)}{2},\quad
F_2 = 1 - \frac{f_+(r)-f_-(r)}{2},\quad
A = 1 - m_e r \tilde a(r).
\eeq
Then we find
\beq
f_+'' + \frac{f_+'}{r}  - m_\lambda^2 f_+ &=& \frac{f_+}{4r^2} - \frac{1}{2r^2} - \frac{3m_\lambda^2}{4}f_+^2  + \frac{m_\lambda^2}{8}f_+^3 
+ {\cal O}(\tilde a^2,\tilde a f_-,f_-^2),\label{eq:f_+}\\
f_-'' + \frac{f_-'}{r}  - m_\eta^2 f_- &=&  \frac{f_-}{4r^2} - \frac{m_e \tilde a}{r} +\left(\frac{m_e \tilde a}{2r} - \left(\frac{m_\lambda^2}{2}+m_\eta^2\right)f_-\right) f_+\non
&&+\frac{1}{4}\left(\frac{m_e \tilde a}{2r} - \left(\frac{m_\lambda^2}{2}+m_\eta^2\right)f_-\right) f_+^2 + {\cal O}(\tilde a^2 f_-,f_-^3),\\
\tilde a'' + \frac{\tilde a'}{r}  - m_e^2 \tilde a &=& \frac{\tilde a}{r^2} - \frac{m_e f_-}{r} - \left(m_e^2 \tilde a - \frac{m_e f_-}{2r}\right) f_+ + \frac{m_e^2}{4}\tilde a f_+^2
+{\cal O}(\tilde a f_-^2).
\eeq
We put  two dimensional static propagators at the left hand side, so that the right hand side are the sources due to the string at the origin.
The masses of the fluctuations $m_+$, $m_-$, $m_e$ for $f_+$, $f_-$, $a$ are consistent with those which can be read from Eq.~(\ref{eq:quadratic_lag}). 
To the leading order, $f_+$ is decoupled with $\tilde a$ and $f_+$.
The first equation has an inhomogeneous source $-1/2r^2$
which appears because of the presence of the global string charge. Thus, the fluctuation $f_+$ decays with
inverse power law as
\beq
f_+ = q_+ K_{\frac{1}{2}}(m_\lambda r) + \frac{1}{2m_\lambda^2 r^2} + {\cal O}\left((m_\lambda r)^{-4}\right),
\eeq
with $K_{\frac{1}{2}}(m r) = \sqrt{\frac{\pi}{2mr}}\,e^{-m r}$ is the modified Bessel function.  
The first term is negligible compared with the second term,
but we keep it because
the subscript $1/2$ of $K_{\frac{1}{2}}(m r)$ is related to a half global string charge.
The leading order term $1/2m_\lambda^2 r^2$ is determined by keeping
the terms up to the linear order in Eq.~(\ref{eq:f_+}), and we should take ${\cal O}(f_+^2,f_+^3)$ into account for obtaining the higher order terms.
Note that Eq.~(\ref{eq:f_+}) also includes the terms of order ${\cal O}(\tilde a^2,\tilde a f_-,f_-^2)$ but they are  exponentially small and are negligible. 
The fluctuations $f_-$ and $\tilde a$ are exponentially small since their equations of motion does not include inhomogeneous source term.
On the other hand, we should not just ignore relatively large factor $f_+$ in the equations of motion for $\tilde a$ and $f_+$.
Generic solution of $f_-$ and $\tilde a$ can be written as
\beq
f_- = q_- K_{\frac{1}{2}}(m_\eta r) + \delta f_-,\qquad
\tilde a = q_a K_1(m_e r) + \delta \tilde a,
\label{eq:asym_fa}
\eeq
where $K_1(m r) = K_0'(mr)/m$ and $K_0(mr) \sim K_1(mr) \sim K_{\frac{1}{2}}(mr)$ at $mr \gg 1$.

When $m_e > m_\eta$ 
we ignore terms of order $e^{-m_e r}$.
Then we find the next leading order correction as
\beq
\delta f_- &=& \left( 
\frac{m_\eta^2 \left(2 m_\eta^2+m_\lambda^2\right)-m_e^2 \left(2 m_\eta^2+5 m_\lambda^2\right)}{8 m_\lambda^2 \left(m_e^2-m_\eta^2\right)}
\frac{1}{m_\eta r}+ {\cal O}(r^{-2}) \right)q_- K_{\frac{1}{2}}(m_\eta r),\\
\delta \tilde a &=& \left(\frac{m_e m_\eta}{m_e^2-m_\eta^2}\frac{1}{m_\eta r}+ {\cal O}(r^{-2})\right) q_- K_{\frac{1}{2}}(m_\eta r).
\eeq

When $m_e < m_\eta$, we ignore term of order $e^{-m_\eta r}$ and find
\beq
\delta f_- &=& \left( \frac{m_e^2}{m_\eta^2-m_e^2} \frac{1}{m_e r} + {\cal O}(r^{-2})\right) q_a K_{\frac{1}{2}}(m_er),\\
\delta \tilde a &=& \left(
\frac{2 m_e^4-m_e^2 \left(2 m_\eta^2+7 m_\lambda^2\right)+3 m_\eta^2 m_\lambda^2}{8 m_\lambda^2 \left(m_\eta^2-m_e^2\right)}
\frac{1}{m_e r}
+ {\cal O}(r^{-2})
\right) q_a K_{\frac{1}{2}}(m_e r).
\eeq

When $m_e = m_\eta$, the leading order in the approximation is $f_- = q_- K_0(m_e r)$ and $\tilde a = q_a K_1(m_er)$.
Hence, we conclude that both $f_-$ and $\tilde a$ have the same asymptotic behavior $f_- \sim \tilde a \sim e^{-m r}$ with $m=\min\{m_e,m_\eta\}$.

\subsection{Stabilizing semilocal string by polarization}

Next, we consider $k_1 = k_2=1$ string.
This can be regarded as the pair of $(k_1,k_2) = (1,0)$ and $(0,1)$ strings.
Although the individual partonic strings have infinite tension due to their global string nature, the $(1,1)$ string has a finite tension because
of the cancelation of the global charges.
This implies the existence an attractive force 
between $(1,0)$ and $(0,1)$ strings at large distance. 
We will see that this is indeed the case if $\eta \neq 0$.
Furthermore, we will also find the string solutions can be coaxial or non-coaxial 
depending on the parameters of the model.
 For any cases,
when we look at the string at sufficiently large distance, the string is almost 
axially symmetric. Therefore, $F  \equiv F_1 \simeq F_2$ asymptotically holds. 
Assuming this relation, the equations of motion reduce to those for the familiar ANO string
\beq
F'' + \frac{F'}{r} - \frac{(1- A)^2}{r^2} F + \frac{m_\lambda^2}{2}\left(1-F^2\right)F \simeq 0,\\
A'' - \frac{A'}{r} + m_e^2 \left(1 - A\right)F^2  \simeq 0.\qquad\qquad
\eeq
The asymptotic behaviors $F = 1 - f$ and $A = 1 - a$ are well known as 
\beq
f = q_s K_0(m_\lambda r) ,\quad
a = q_a K_1(m_e r).
\eeq
The mass parameter $m_\eta$ does not contribute to determine the asymptotic behaviors in this case,
but should affect some local substructure of the semilocal string solutions as we will see below.

The semilocal strings in this model have many interesting features. 
Among them, the most interesting one is stabilization of the semilocal string in the type II region ($m_e < m_\lambda$) by finite polarization.

Let us start with defining polarization of the semilocal string. As before, let $z_1$ and $z_2$ be zeros of $H_1$ and $H_2$, respectively.
Then the polarization is defined by
\beq
P =\frac{d}{2},\quad d \equiv z_1 - z_2,
\label{eq:polarization}
\eeq
where $d$ stands for the displacement vector. We put the factor $1/2$ because the semilocal string can be interpreted as a pair of fractional strings.
In order to illustrate the situation, consider a configuration given by
\beq
H_1 = v \frac{z-z_1}{\sqrt{|z-z_1|^2 + |z-z_2|^2}},\quad
H_2 = v \frac{z-z_2}{\sqrt{|z-z_1|^2 + |z-z_2|^2}}.
\label{eq:lump}
\eeq
Note that these satisfy $|H_1|^2 = |H_2|^2 = v^2/2$.
Then, consider a closed contour $C_1$ encircling only $z=z_1$. When we go around $z=z_1$ along $C_1$, the phase of $H_1$ changes by $2\pi$
while the phase of $H_2$ remains intact. This is possible because we have two $U(1)$ symmetries: the one is the gauge symmetry and the other
is global symmetry. The $U(1)_{\rm gauge} \times U(1)_{\rm global}$ charges for the configuration like Eq.~(\ref{eq:lump}) 
are $(+1,+1)$ for $H_1$ and $(+1,-1)$ for $H_2$, respectively.
Thus, when we go around $z=z_1$, we travel around $U(1)_{\rm gauge}$ by $+\pi$ and $U(1)_{\rm global}$ by $+\pi$.
On the other hand, when we go around $z=z_2$, we travel around $U(1)_{\rm gauge}$ by $+\pi$ and $U(1)_{\rm global}$ by $-\pi$. 
Namely, $H_1$ has a half quantized winding number $(\frac{1}{2},\frac{1}{2})$ while
$H_2$ has a half quantized winding number $(\frac{1}{2},-\frac{1}{2})$. 
Thus,  
a single semilocal string carries the topological charge $+1$ for $U(1)_{\rm gauge}$ 
and no charges for  $U(1)_{\rm global}$,
but each fractions$\pm \frac{1}{2}$ fractional winding numbers for $U(1)_{\rm global}$.
This is the reason why we put factor $1/2$ in Eq.~(\ref{eq:polarization}).
The polarization $P$ is indeed the dipole moment for the $U(1)_{\rm global}$ 
topological charge.

Note that the polarization for the semilocal string is quite obscure at $\eta = 0$. This is because $U(1)_{\rm global}$ 
is a subgroup of the manifest $SU(2)$ flavor symmetry. In the $SU(2)$ symmetric model, as is shown in Eq.~(\ref{eq:zero_shift}),
the zeros of Higgs fields can be moved by the $SU(2)$ transformation. Accordingly, the configuration is perfectly axially symmetric, so 
that the semilocal strings in the model with $\eta = 0$ have no dipole-like behaviors.
Nevertheless, as long as we restrict ourselves to the solutions with the fixed boundary condition $|H_1| = |H_2| = v/\sqrt{2}$,
the term ``polarization'', which is usually called ``size'' in the literature, is still useful.
In this sense, we would say the semilocal string at $\eta=0$ is unpolarized for the type I region, and {\it infinitely} polarized for the type II region.
At the critical point (BPS), the polarization can be freely changed.

\begin{figure}[t]
\begin{center}
\includegraphics[width=17cm]{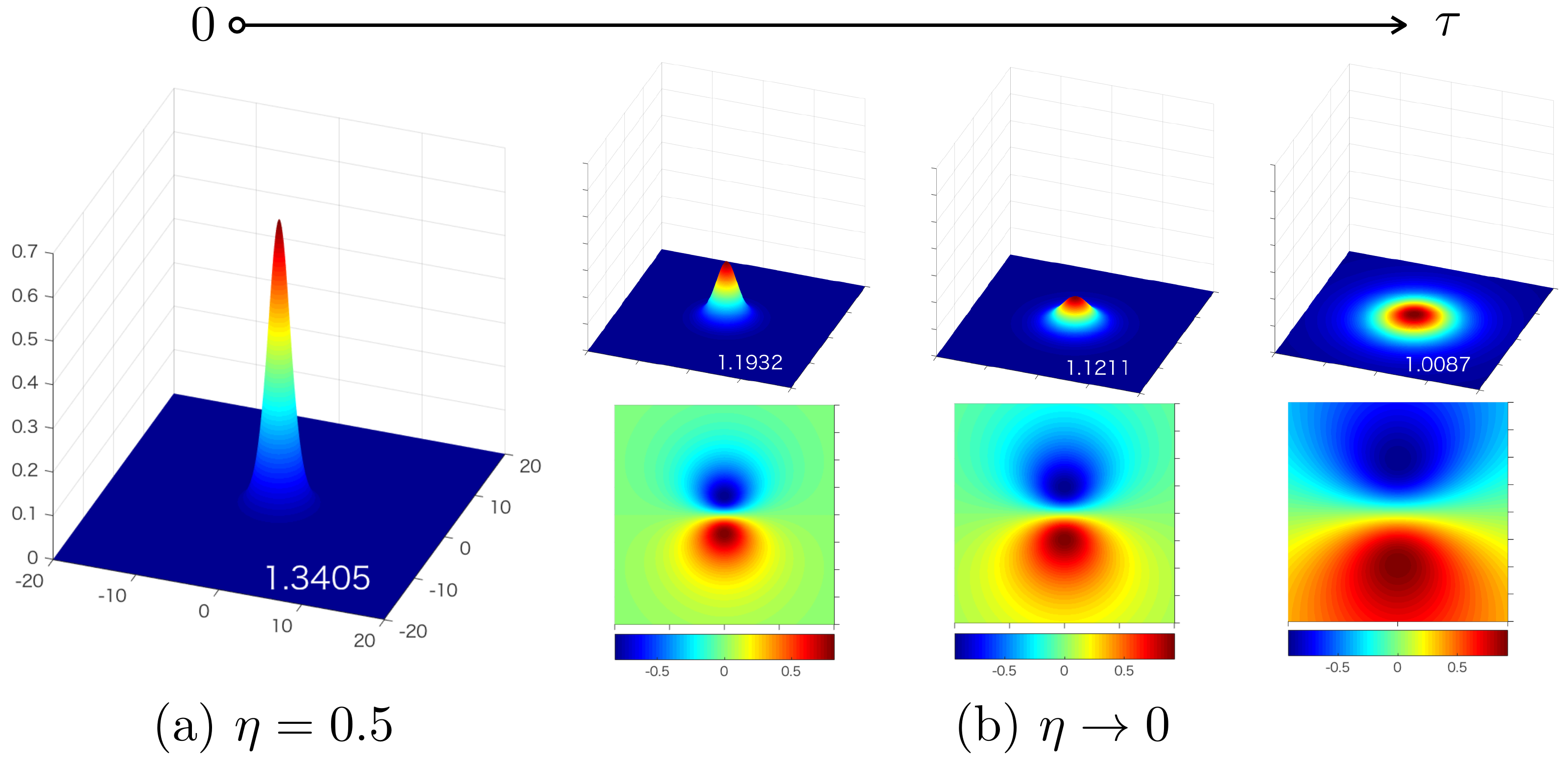}
\caption{(a) Energy density plot of the stable semilocal string in the type IIb with $(e,\lambda,\eta) = (0.3,0.6,0.5)$ and $v=1$. 
The string is axially symmetric and is unpolarized because
$|H_1| = |H_2|$ everywhere. (b) Snapshots of the 2nd energy relaxation evolution for $\eta = 0$ 
with taking the configuration in the panel (a) as an initial configuration at $\tau = 0$. The upper panels show the energy densities 
(the axes are the same as those of (a))
and the lower panels show $H\tau_3H^\dagger$.
The numerical values plotted in the upper panels are the tension in the unit of $2\pi v^2$.}
\label{fig:instability}
\end{center}
\end{figure}

Let us now come back to the semilocal string in the type II region for the $\eta \neq 0$ case.
The first example is the semilocal string with $(e,\lambda,\eta) = (0.3,0.6,0.5)$ which is in the type IIb ($m_e < m_\eta < m_\lambda$).
We find the axially symmetric semilocal string, see the panel (a) in Fig.~\ref{fig:instability}. For this solution, we find
$|H_1| = |H_2|$ everywhere on the $xy$ plane, so that it is unpolarized. The additional Higgs potential $V_1$ forces $|H_1|$ and $|H_2|$ to be equal, and
therefore the number of flavor is essentially one. Namely, the semilocal string solution is precisely the same as the type II ANO string.
Indeed, if we impose $H_1 = H_2$, the terms related to $V_1$ in the equations of motion vanish and we are left with the equations of motion for the ANO string 
in the Abelian-Higgs model with one complex scalar. 
The numerical solution shown in the panel (a) in Fig.~\ref{fig:instability} is obtained by the energy relaxation process explained in Appendix~\ref{sec:NR}.
If we further try to sweep out energy by continuing the imaginary time evolution, we find that changes are
pretty tiny, which proves the solution already reaches the convergent point.
Now, let us see what will happen as $\tau$ goes by after we suddenly turn off $\eta \to 0$. Namely, we perform the 2nd relaxation process with
the configuration shown in Fig.~\ref{fig:instability}(a) for $\eta = 0$. The model with $\eta = 0$ is in the type II region $(e,\lambda,\eta) = (0.3,0.6,0)$ of the
$SU(2)$ symmetric model. Therefore, the initial configuration is no longer stable and dilutes to the vacuum, see also the effective potential given
in Fig.~\ref{fig:effpot_size_moduli}.
In the panels in Fig.~\ref{fig:instability}(b), we show several snapshots for the transition. The semilocal vortex expands with axially symmetric shape being kept.
The plots in the bottom line of Fig.~\ref{fig:instability}(b) shows $H\tau_3H^\dagger$ for each moment.
The darkest red point corresponds to zero of $H_1$ while the darkest blue point to zero of $H_2$ ($H\tau_3H^\dagger$ is initially zero everywhere).
It is clearly seen that the dilution of the type II semilocal string in the $SU(2)$ symmetric model is accompanied by $|z_1 - z_2| \to \infty$ as expected.
Thus, the role of the additional potential $V_1$ is to prevent $d = z_1-z_2$ from flying to infinity. Indeed, the additional Higgs potential energetically prefer 
$H_1 = H_2$ to $H_1 \neq H_2$. This is reason why the semilocal string does not expand for $\eta \neq 0$.

\begin{figure}[t]
\begin{center}
\includegraphics[width=16cm]{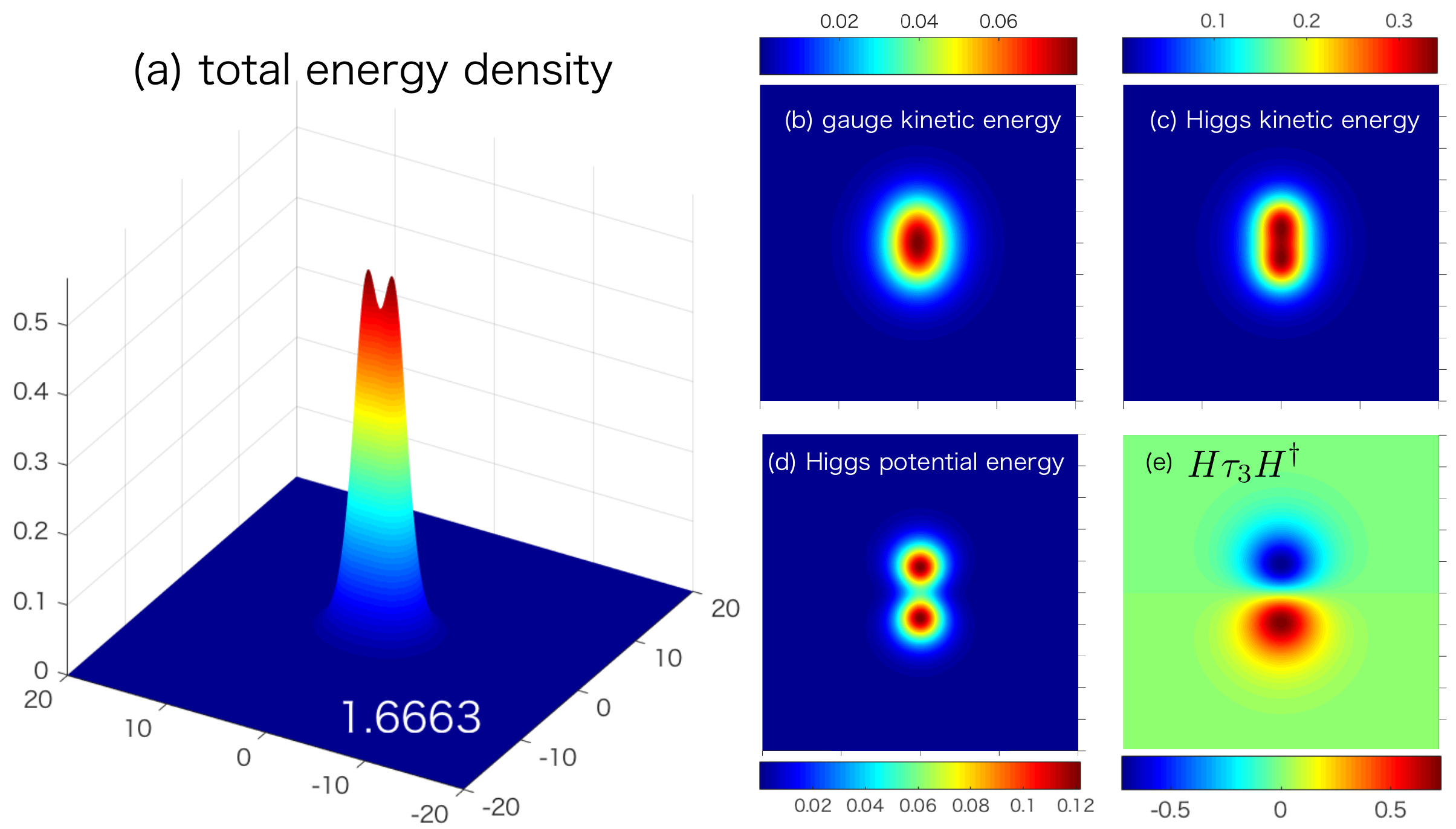}
\caption{Energy densities of the single semilocal string with two partons. The parameters are $v=1$ and $(e,\lambda,\eta)=(0.3,1.2,0.5)$. 
The panels (a) -- (e) show the total energy density, the gauge kinetic energy $F_{12}^2/2e^2$,
the Higgs kinetic energy $\sum_a|\D_i H_a|^2$, the Higgs potential energy $V_0+V_1$, and $H\tau_3H^\dagger$, respectively.
The total energy density peaks are located at $(\pm 1.2/v,0)$ and the Higgs zeros are at $(\pm 1.7/v,0)$. The plot regions for the four small panels are $[-10,10]^2$.
The number at the corner in (a) is the tension in the unit of $2\pi v^2$.}
\label{fig:typical_conf}
\end{center}
\end{figure}

For the above parameter choice $(e,\lambda,\eta) = (0.3,0.6,0.5)$, the effect of $V_1$ is relatively too strong.
As a consequence,  $z_1$ is stuck to $z_2$ and the semilocal string is unpolarized. This solution is less interesting because
it is precisely same as the type II ANO string in the Abelian-Higgs model 
with a single Higgs field.
One may expect that if we weaken the effect of $V_1$, the displacement $d = z_1 - z_2$ is neither infinity nor zero.
It is indeed the case. As a typical example, let us take $(e,\lambda,\eta) = (0.3,1.2,0.5)$.  
The value $\lambda=1.2$ is twice bigger than the previous one. The corresponding semilocal
string solution is shown in Fig.~\ref{fig:typical_conf}.  As is clearly shown, the configuration is not axially symmetric, and the total energy density
shown in Fig.~\ref{fig:typical_conf}(a) has two peaks at $(x,y) = (\pm 1.2/v,0)$. The total energy density is the sum of the gauge kinetic contribution in Fig.~\ref{fig:typical_conf}(b),
the Higgs kinetic contribution in Fig.~\ref{fig:typical_conf}(c), and the Higgs potential energy in Fig.~\ref{fig:typical_conf}(d).
The gauge kinetic energy is elliptic shape with a single peak at the origin, whereas the Higgs potential has sharp two peaks whose locations are almost 
coincide with the Higgs zeros $(x,y) = (\pm 1.7/v,0)$ which are identical to the positions of positive and negative peaks of $H\tau_3H^\dagger$ shown
in Fig.~\ref{fig:typical_conf}(e).
\begin{figure}[t]
\begin{center}
\includegraphics[width=16cm]{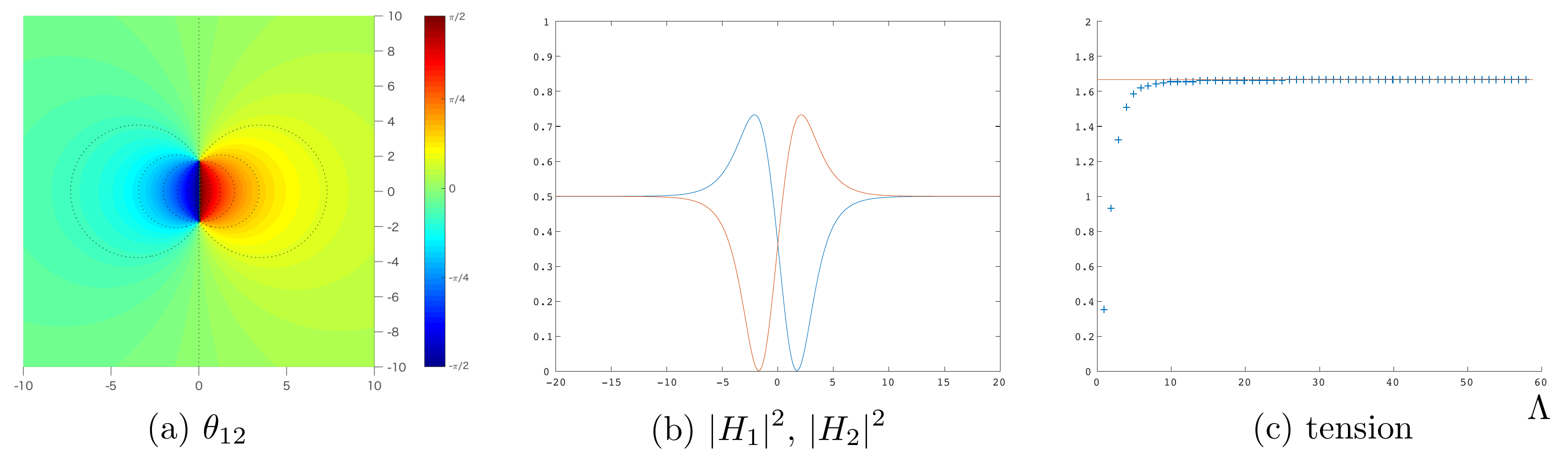}
\caption{(a) The relative phase $\theta_{12}$ in the $xy$ plane for the semilocal string with $(e,\lambda,\eta)=(0.3,1.2,0.5)$.
(b) The profile funtions of $H_1$ (red) and $H_2$ (blue). (c) The tension in the unit of $2\pi v^2$
as a function of IR cutoff $\Lambda$. The red line shows the asymptotic value $1.6663$ and the blue crosses are numerically
obtained value of $T(\Lambda)$.}
\label{fig:relative_phase}
\end{center}
\end{figure}
In order to see the expected dipole structure in this solution, 
it is worth to plot the relative phase 
\beq
\theta_{12} = \frac{1}{2}\arg(H_1/H_2)
\eeq
as shown in Fig.~\ref{fig:relative_phase}. 
There is a branch cut between $(x,y) = (-1.7/v,0)$ and $(1.7/v,0)$.
When we go around the lower/upper branch point, $\theta_{12}$ changes by $+\pi$/$-\pi$.
Namely, the semilocal string is a dipole of the half quantized global $U(1)$ charges. 
The polarization is given by
\beq
P = \frac{1.7}{2v}.
\eeq
We also show the amplitudes of the Higgs fields in Fig.~\ref{fig:relative_phase}(b), in which one can clearly see $H_1$ and $H_2$ have
zeros at different points. Finally, in order to check if the tension is finite, we plot $T(\Lambda) = \int^{\Lambda}_0dr\int^{2\pi}_0 d\theta\, r {\cal H}$
in Fig.~\ref{fig:relative_phase}(c).  As shown in the figure, $T(\Lambda)$ rapidly converges to a finite value. This is sharp contrast to $[\frac{1}{2},\frac{1}{2}]$ string
solution shown in Fig.~\ref{fig:local_global_vortex}, where $T(\lambda)$ is logarithmically divergent. The string solution here has $[1,0]
= [\frac{1}{2},\frac{1}{2}] + [\frac{1}{2},-\frac{1}{2}]$ has zero net global charge, so that the tension remains finite as usual local strings.

\subsection{Phase diagram}

In this subsection, we will survey the parameter space $(e,\lambda,\eta)$.  
Especially, we are interested in clarifying when the semilocal vortex is polarized.
Before doing this, let us remind that the usual ANO string in the Abelian-Higgs model 
with one complex scalar field is essentially controlled by
one dimensionless parameter 
\beq
\gamma = \frac{m_e}{m_\lambda}.
\eeq
The ANO string solutions with different $(m_e,m_\lambda)$ and $(m_e',m_\lambda')$ 
are essentially the same solution up to overall coordinate rescaling
if  $m_e/m_\lambda = m_e'/m_\lambda'$ holds. This is the reason why the ANO
strings are classified into three types: 
the type I $(\gamma > 1)$, type II $(\gamma < 1)$ and the critical (BPS) $(\gamma = 1)$.
In our model, we have four different masses $0$, $m_e$, $m_\lambda$ and $m_\eta$. Therefore, classification of the strings in our model is
more complicated. For example, even in the case of $m_e/m_\lambda = m_e'/m_\lambda'$, the corresponding string solutions might essentially be different. 

\begin{figure}[t]
\begin{center}
\includegraphics[width=16cm]{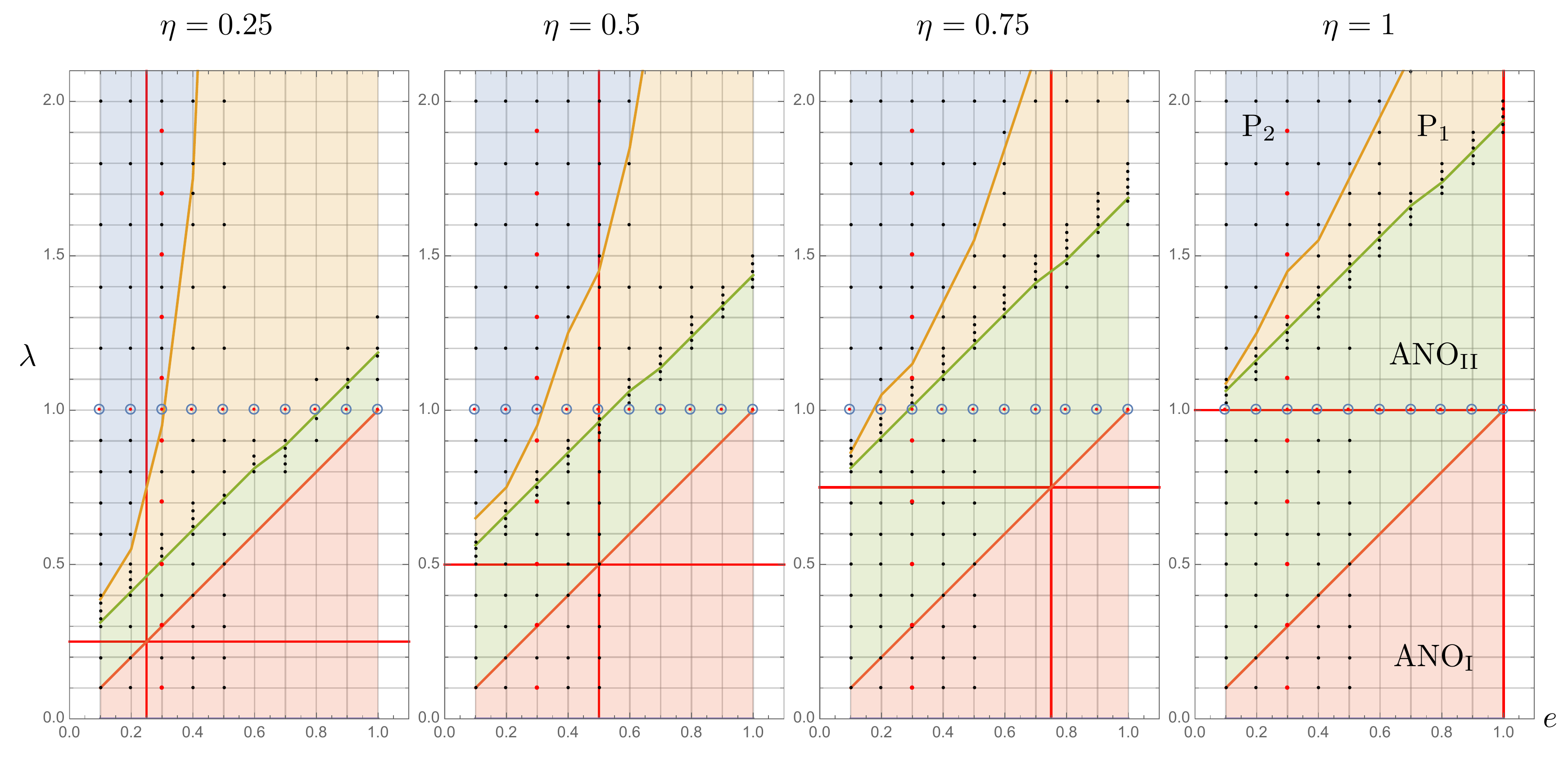}
\caption{Phases of the $(k_1,k_2) = (1,1)$ string in the $e$-$\lambda$ plane with fixed values of $\eta = 0.25, 0.5, 0.75, 1$.
The four regions, ANO$_{\rm I}$, ANO$_{\rm II}$, $P_1$ and $P_2$, correspond to pink, green, yellow and blue regions, respectively.
ANO$_{\rm C}$ corresponds to the points on the line $e=\lambda$.
The parameters for which we obtained
numerical solutions are expressed by the markers on cites.
The red and orange lines correspond to the dashed lines in Fig.~\ref{fig:elambda} dividing the parameter space into six regions.
}
\label{fig:phase_diagram}
\end{center}
\end{figure}

We classify the solutions into 2 categories: The ANO type or polarized (P) (or molecular) type 
according to the displacement $d=0$ or $d\neq 0$, respectively. 
(We refer the ANO string by an unpolarized string.)
Looking into the details, the ANO type can further be classified into ANO$_{\rm I}$, ANO$_{\rm II}$ and ANO$_{\rm C}$ according to $m_\lambda < m_e$,
$m_e < m_\lambda$, and $m_e = m_\lambda$. 
Similarly, the P type is decomposed into P$_1$ (P$_2$) for configuration with one peak (two peaks) in the total energy density.
For example, the configuration given in Fig.~\ref{fig:instability}(a) is of the type ANO$_{\rm II}$, and the solution in Fig.~\ref{fig:typical_conf} is of the type P$_2$.
In Fig.~\ref{fig:phase_diagram}, we show the $e$-$\lambda$ plane with fixed values of $\eta = 0.25, 0.5, 0.75, 1$. 
 The parameters for which we really obtained
numerical solutions are expressed by the markers on cites in Fig.~\ref{fig:phase_diagram}.
Observing the phase diagram in Fig.~\ref{fig:phase_diagram}, we soon realize that there are stable strings 
in the type II region ($e < \lambda$) where the usual semilocal string at $\eta = 0$ is unstable.
Indeed, we find finite energy and finite size string solutions for all $|\eta| > 0$. 
Namely, the type II semilocal strings are stabilized as long as
$\eta$ is nonzero, in contrast to the common knowledge that 
type-II semilocal strings should be unstable.

Let us see how the string configuration changes as varying the parameters. 
\begin{sidewaysfigure}
\includegraphics[width=\columnwidth]{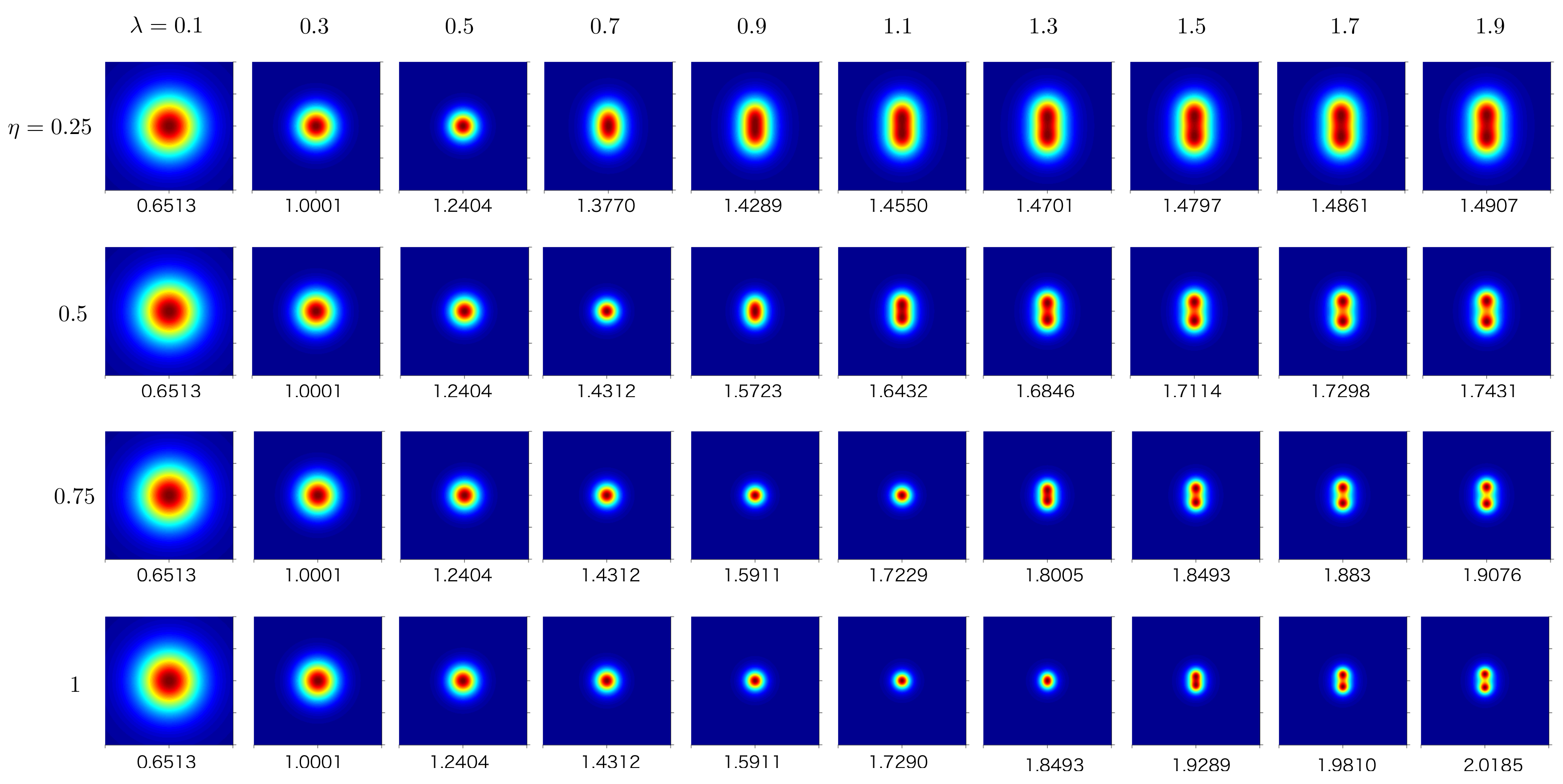}
\caption{Numerical solutions for $\lambda$-$\eta$ plane with $e=0.3$.
The numerical values below the figures are the tension in the unit of $2\pi v^2$.
}
\label{fig:conf_lambda_eta}
\end{sidewaysfigure}
\begin{sidewaysfigure}
\includegraphics[width=\columnwidth]{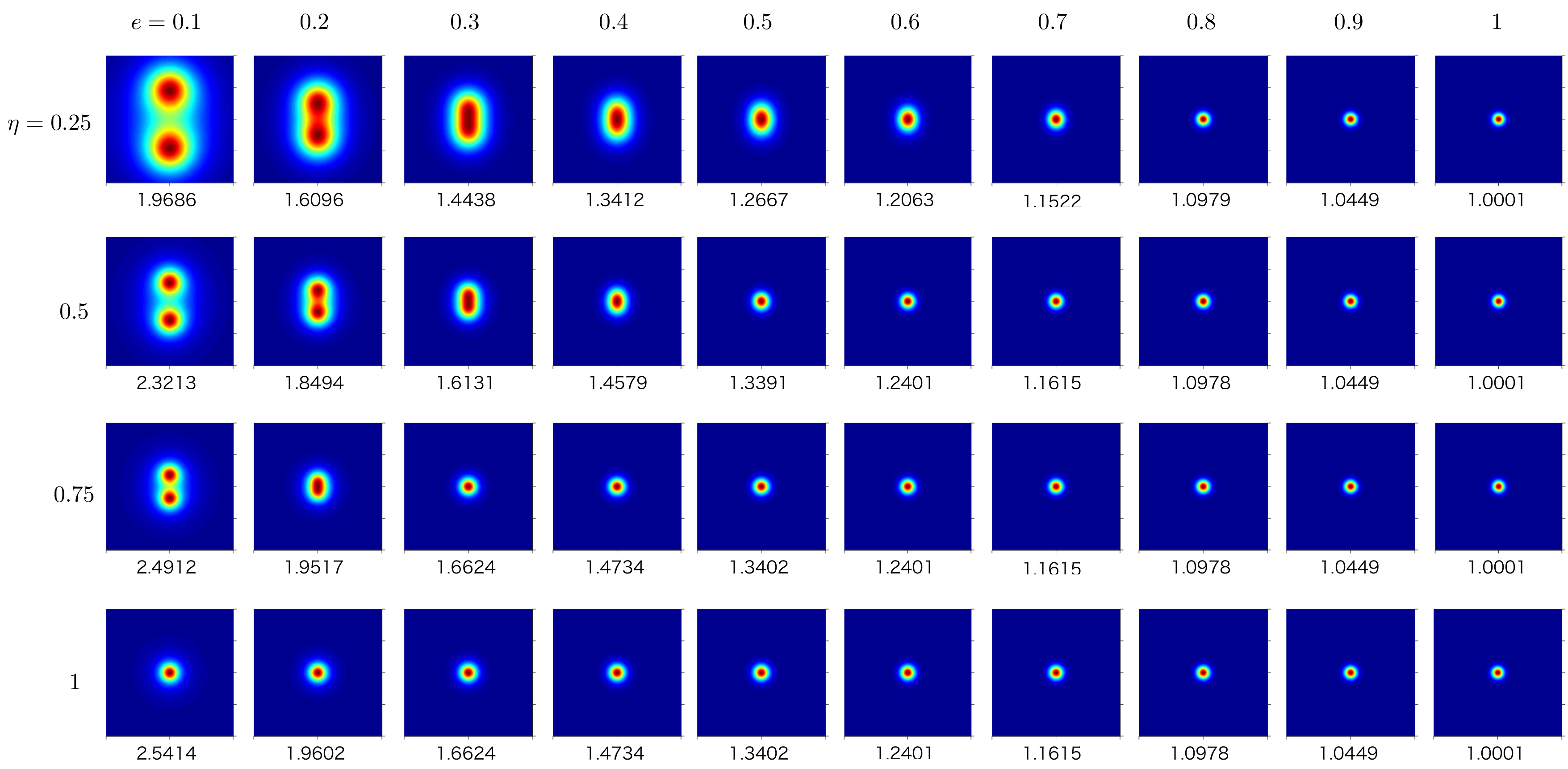}
\caption{Numerical solutions for $e$-$\eta$ plane with $\lambda=1$.
The numerical values below the figures are the tension in the unit of $2\pi v^2$.}
\label{fig:conf_e_eta}
\end{sidewaysfigure}

\begin{figure}[t]
\begin{center}
\includegraphics[width=16cm]{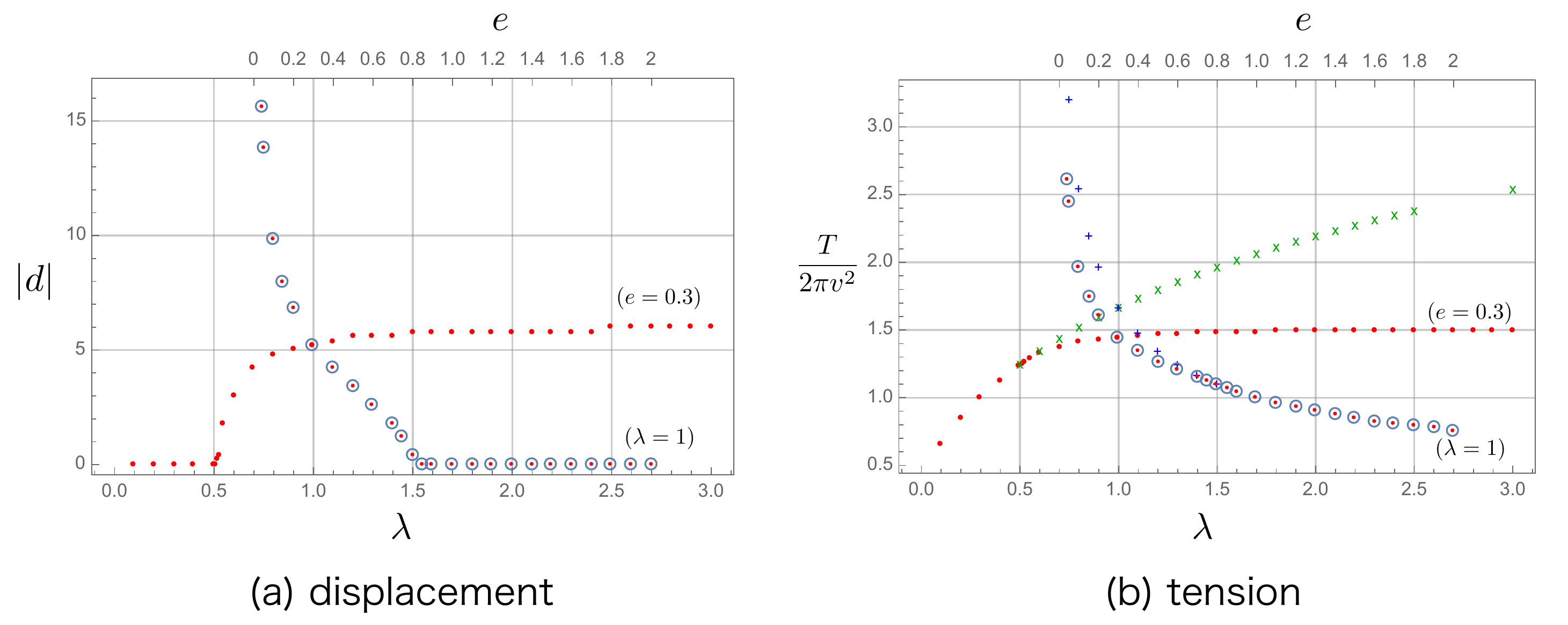}
\caption{Displacement (a) and tension (b) as functions of $e$ and $\lambda$ for $\eta = 0.25$.
The corresponding solutions (the red-dotted cites and the red-dotted cites in the blue circle) 
are shown in the left-most panel of Fig.~\ref{fig:phase_diagram} and
the subfigures in the top columns of Figs.~\ref{fig:conf_lambda_eta} and \ref{fig:conf_e_eta}.
In the panel (b), the green crosses (blue crosses) show the tension as a function of $\lambda$ ($e$) for the ANO solutions in the $\eta=0$ limit.
}
\label{fig:displacement}
\end{center}
\end{figure}

For concreteness, first, we change $\lambda$ with $e$ and $\eta$ being fixed, namely move
vertically in Fig.~\ref{fig:phase_diagram}. 
Let us focus on the strings with $e=0.3$ which correspond to the red-dotted cites in each panels of Fig.~\ref{fig:phase_diagram}. 
The corresponding energy densities for the red cites are shown in Fig.~\ref{fig:conf_lambda_eta}.
The strings with $\lambda < e$ in the left-most column are of the ANO$_{\rm I}$ type and those in the second column from left are of the BPS type.
Increasing  $\lambda$ beyond the BPS line $(e=\lambda)$, the strings now enter the ANO$_{\rm II}$ region. All these strings are unpolarized because
the Higgs fields are forced to be $H_1 = H_2$ at any spacetime points. 
Departing further from BPS line toward larger $\lambda$, 
now the configurations transit to be polarized. 
Namely, $H_1$ and $H_2$ have zeros at different points. 
The string cross section is elongated, so that shape of the total energy density
becomes elliptic with single peak (P$_1$) for relatively small $\lambda$ as can be seen 
for example $\lambda = 0.7,\,0.9$ in
the top line of Fig.~\ref{fig:conf_lambda_eta}, or acquires
two peaks (P$_2$) for sufficiently large $\lambda$ as can be seen in 
the panels with $\lambda \ge 1.1$ in the top line of Fig.~\ref{fig:conf_lambda_eta}.
The relation between the displacement $|d|$ and $\lambda$ for $e=0.3$ and $\eta = 0.25$ is shown in Fig.~\ref{fig:displacement}.
Universal feature for changing $\lambda$ under fixing $e$ is that $d$ is zero below  a certain $\lambda_0(\eta)$, and
$|d|$ increases above $\lambda_0(\eta)$. It seems that $|d|$ converges to an upper value $|d_{\rm max}(\eta)|$ for $\lambda \to \infty$,
see Fig.~\ref{fig:displacement}(a).
The critical value $\lambda_0(\eta)$ as a function of $\eta$ is a monotonically increasing function. 
We also find that $|d_{\rm max}(\eta)|$ is a monotonically decreasing
function of $\eta$.

Next, we fix $\lambda$ and $\eta$, and vary $e$. It is horizontal movement in 
the $e$-$\lambda$ plane in Fig.~\ref{fig:phase_diagram}.
As typical configurations, we look at the strings with $\lambda = 1$ which correspond to the red-dotted cites inside blue circle in
Fig.~\ref{fig:phase_diagram}.
The corresponding total energy density are shown  in Fig.~\ref{fig:conf_e_eta}.
The displacement $|d|$ tends to be larger for smaller $e$ while it becomes zero for $e$ above 
a certain critical $e_0(\eta)$.
The detailed relation between $|d|$ and $e$ is shown in Fig.~\ref{fig:displacement}(a). 
As opposed to large $\lambda$ limit, $|d|$ becomes steeply large at $e$ vanishing limit.
The critical value $e_0(\eta)$ as a function of $\eta$ is a monotonically decreasing function. 

Finally, let us see what happens when we vary $\eta$ with $e$ and $\lambda$ being fixed.
As can be seen in Figs.~\ref{fig:conf_lambda_eta} and \ref{fig:conf_e_eta}, the displacement $|d|$ becomes small when we increase $\eta$.
However, once the string becomes unpolarized, the configuration is frozen. 
Namely, it is not further deformed even if we further increase $\eta$.

From above observations, we find an inclination for $|d|$ to become larger
when we come deeper into the type II region $(e \ll \lambda)$. This behavior may be expected if we recall the well-known fact 
that two integer ANO strings repel in the type II region as is already mentioned. 
However, our strings are not integer ANO strings but
two strings with fractional local and global charges. 
Indeed, we find that both two limits $e \to 0$ and $\lambda \to \infty$ with $\eta$ being fixed lead to the polarization, but 
they are qualitatively different. The displacement $d$ diverges at the former limit while it converges to a finite value
in the latter limit.
From numerical observation in Fig.~\ref{fig:phase_diagram}, we have an estimation for the critical values $\lambda_0(\eta)$ and $e_0(\eta)$ as functions of $\eta$ as
\beq
\lambda_0(\eta) - e_0(\eta) \simeq \eta.
\eeq 
This equation explains the existence of the unpolarized ANO$_{\rm II}$ string in the band $e < \lambda \lsim e+\eta$
(the light-green region in Fig.~\ref{fig:phase_diagram}).
The polarized P$_{1,2}$ string appears in the region $\lambda \gsim e + \eta$ which is above the ANO$_{\rm II}$-band
(the light-blue and light-orange regions in Fig.~\ref{fig:phase_diagram}).
Note that the ANO$_{\rm II}$-band closes when $\eta= 0$, and the polarized strings become unstable because their  polarizations become infinite.
In terms of the classification by (IIa/IIb/IIc) as given in Fig.~\ref{fig:elambda}, 
we find that the only ANO$_{\rm II}$ string 
appears in the IIa region. Both the ANO$_{\rm II}$ and $P_{1,2}$ are possible in the IIb and IIc  regions, 
but the dominant part of 
the IIb region is occupied by P$_2$ strings while the IIc region is dominated by P$_1$ strings.

\section{Summary and discussion} \label{sec:summary}

The semilocal strings exist in the extended Abelian-Higgs model with 
the two complex scalars with the $SU(2)$ symmetry.
It has been known that the semilocal strings are stable only in the type I parameter region $m_e > m_\lambda$ but unstable to expand in the type-II parameter region $m_e < m_\lambda$ .
In this paper, we have studied stabilization of the semilocal strings by the polarization in the presence
of the additional $SU(2)$ breaking potential given in Eq.~(\ref{eq:add_pot}).  
We have found that the semilocal strings in the type II region ($m_e < m_\lambda$)
become stable against expansion by the additional potential. 
The single semilocal string splits into two fractional strings with
opposite global charges $\pm 1/2$. Thus, the semilocal string is polarized.
It is crucial that the non-Abelian flavor symmetry is broken by the
additional potential, which allows the existence of fractionally charged string with $1/2$ local and $1/2$ global charge.
We have obtained the numerical solutions for the fractional strings 
for various parameters.
We also have investigated the asymptotic behaviors 
and have found that they decay exponentially with the smallest masses of the fields at the bulk, 
which are quite different from those of the well-known ANO strings and usual semilocal strings.
We further have studied dependence of the polarization of the single semilocal string on the masses $m_e$, $m_\lambda$ and $m_\eta$ in detail.
We have found that the semilocal string is stable in the whole parameter region. 
Especially, the type II region $m_e < m_\lambda$ is divided into
two phases. In one of them, 
the unpolarized semilocal strings, namely the type-II ANO solutions, 
appear for $m_e < m_\lambda < m_e + m_\eta$.
In the other region $m_\lambda > m_e + m_\eta$, the two Higgs fields have zeros at different points, namely the strings are polarized.
The displacement $|d|$ of the two zeros is larger for smaller $m_e$. It also increases as 
$m_\lambda$ is increased, but it saturates some upper value.

Before closing this paper,
several discussions are addressed here.

We have studied only a single vortex in this paper.
Multiple vortices including the interaction among them are 
an important next step.  
In particular, two neighboring vortices will enhance their polarizations 
because of attraction between fractional vortices belonging 
to each other.  
For a small number of vortices, vortex molecules may constitute 
a vortex polygon, that is vortices sit at the vertices of a polygon, 
as the case of two-component BECs \cite{Kobayashi:2013wra}. 
For a large number of vortices, we may expect that they 
constitute a vortex lattice when the system size is finite,
as the case of conventional type-II superconductors 
in the presence of an applied magnetic field. 
In the $SU(2)$ symmetric extended Abelian Higgs model of $\eta=0$, 
a vortex lattice will be difficult to be realized from the following reason.
Vortices are unstable in the type-II region 
and stable in the type-I region.
However, in the type-I region, 
vortices are attractive and so superconductors 
are unstable against the applied magnetic field. 
As shown in this paper, in the presence of the additional potential term 
($\eta \neq 0$),
semilocal vortices become stable even in the type-II region.
Consequently, 
these superconductors are stable against
the applied magnetic field,
in which a vortex lattice will be formed.
The form of vortex lattice may be similar to 
that of two-component BECs under the rotation \cite{Mueller:2002,Kasamatsu:2003,Kasamatsu:2004,Kasamatsu:2005,Aftalion:2012,Cipriani:2013nya}
or that of two-gap superconductors  
where vortices are polarized
\cite{fractional-exp}. 

If we consider a potential term
\beq
V_2 = \frac{\eta^2}{2} (H\tau_3 H^\dagger - r^2)^2 
= \frac{\eta^2}{2} (|H_1|^2 - |H_2|^2 - r^2)^2,
\label{eq:add_pot2}
\eeq
instead of Eq.~(\ref{eq:add_pot2}) considered in this paper, 
the fluxes of fractional vortices deviate from 1/2.
This phenomenon is known for multi-gap superconductors, multi-component BECs, 
and ${\mathbb C}P^1$ lumps. 

In addition to the potential term in Eq.~(\ref{eq:add_pot2}) that we considered in this paper,
we may further add the potential term $\tr (H \tau_2 H^\dagger)$.
This is known as an intrinsic Josephson interaction term.
In this case, 
two fractional vortices constituting a single semilocal vortex 
will be connected by a sine-Gordon kink,
as the case of two-gap superconductors \cite{Tanaka:2001,Goryo:2007}
or coherently coupled two-component BECs \cite{Son:2001td,Cipriani:2013nya}.

In this paper, we have studied the extended Abelian-Higgs model 
with two Higgs fields.
Generalization to $N$ Higgs fields is possible,
where semilocal vortices reduce to 
${\mathbb C}P^{N-1}$ sigma model lumps 
in the strong gauge coupling limit.
In this case, we may add a potential term
\beq
V_2 = \sum_a {\eta_a^2\over 2} \tr (H h_a H^\dagger)^2
\eeq 
where $h_a$ are all possible Cartan generators of $SU(N)$.
Then, one semilocal vortex is split into $N$ fractional vortices 
with $1/N$ quantized fluxes.
Further adding Josephson terms
$\tr (H E_{ij} H^\dagger)$ with $E_{ij}$ having 
a nonzero $(i,j)$ component 
is also interesting, 
by which $i$-th and $j$-th fractional vortices are connected.
The total configuration would form a vortex graph, 
as the case of multi-component BECs
\cite{Eto:2012rc,Eto:2013spa}.

\section*{Acknowledgments}

This work is supported by the Ministry of Education,
Culture, Sports, Science (MEXT)-Supported Program for the Strategic
Research Foundation at Private Universities ``Topological Science''
(Grant No.~S1511006), and  the Japan Society for the Promotion of Science
(JSPS) Grant-in-Aid for Scientific Research (KAKENHI Grant
No.~25400268).
The work of M. E. is supported in part by
JSPS Grant-in-Aid for Scientific Research (KAKENHI Grant No.~26800119).
The work of M.~N.~is supported in part by a Grant-in-Aid for
Scientific Research on Innovative Areas ``Topological Materials
Science'' (KAKENHI Grant No.~15H05855) and ``Nuclear Matter in Neutron
Stars Investigated by Experiments and Astronomical Observations''
(KAKENHI Grant No.~15H00841) from the MEXT of Japan.

\begin{appendix}
\section{Numerical recipe}
\label{sec:NR}

We numerically solve  the equations of motion (\ref{eq:eom0_2}), (\ref{eq:eom0_1_mod}) and (\ref{eq:eom0_2_mod}) under
a static assumption as $A_{1,2}(x^1,x^2)$ and $H(x^1,x^2)$. We make an ansatz $A_0 = A_3 = 0$, and
take a gauge $\p_1 A_1 + \p_2 A_2 = 0$ throughout this paper.
Instead of solving directly the equations of motion for the fileds $X=\{A_1,A_2,H_a\}$, 
we will solve the following gradient flow equations 
\be
\p_i^2 X(x^1,x^2,\tau) + {\cal U}(X(x^1,x^2,\tau)) = \p_\tau X(x^1,x^2,\tau),
\label{eq:GF}
\ee 
where an ideal time $\tau$-dependence is introduced. The original equations of motion are obtained by
letting the right hand side to be zero. With an appropriate initial function $X_0 = X(x^1,x^2,\tau =0)$ which has qualitatively the same
behaviors as Eq.~(\ref{eq:asymp}), we solve
the time evolution of $X(x^1,x^2,\tau)$. Typically, $\p_\tau X$ gradually goes to zero as the time evolution.
As a consequence, we get the solution $X(x^1,x^2,\tau = \infty)$ to the original equation of motion.
Our computational box is typically $[-60,60]^2$ divided into $1200^2$ lattice points,
and we will solve the gradient flow equation (\ref{eq:GF}) by the Crank-Nicolson type method.
We take the Neumann boundary conditions for all the fields.

In the following, we will set $v=1$. In other words, we will use rescaled variables as
\be
\tilde x^\mu = v x^\mu,\quad
\tilde A_\mu = A_\mu / v,\quad
\tilde H = H/v.
\ee
Then, $v$ dependence disappears from Eqs.~(\ref{eq:eom0_2}) and (\ref{eq:eom0_1_mod}).
We will not distinguish $X$ and $\tilde X$ unless stated otherwise.
\end{appendix}

\end{document}